\documentclass{article}
\usepackage{graphicx}	
\usepackage{subfig}	
\usepackage{amssymb}
\usepackage[letterpaper]{geometry}
\usepackage[export]{adjustbox} 
\usepackage{tikz}

\usepackage{amsmath,amssymb,amsthm}
\usepackage{bm}

\newlength{\figwidth}
\newlength{\figwidthtwo}
\newlength{\figwidththree}
\newcommand{\fref}[1]{Fig.\,\ref{#1}}
\newcommand{\tref}[1]{Table\,\ref{#1}}
\newcommand{\eref}[1]{Eq.\,(\ref{#1})}

\newcommand{\sref}[1]{Sec.\!~\ref{#1}}

\newcommand{\cref}[1]{Ref.\,\cite{#1}}
\newcommand{\crefs}[1]{Refs.\,\cite{#1}}
\newcommand{\fnote}[1]{\footnote{#1}}

\newcommand{\vs}{{\it vs.}\! }
\newcommand{\ie}{{\it i.e.}\! }
\newcommand{\eg}{{\it e.g.}\! }
\newcommand{\etc}{{\it etc.}\! }
\newcommand{\etal}{{\it et al.}\! }

\newcommand{\stress}{{\sigma}}

\newcommand{\pluseq}{\mathrel{+}=}

\newcommand{\as}{\mathsf{a}}
\newcommand{\bs}{\mathsf{b}}
\newcommand{\ds}{\mathsf{d}}
\newcommand{\xs}{\mathsf{x}}
\newcommand{\ys}{\mathsf{y}}
\newcommand{\Ws}{\mathsf{W}}
\newcommand{\Ac}{\mathcal{A}}
\newcommand{\Bc}{\mathcal{B}}
\newcommand{\Ic}{\mathcal{I}}

\newcommand{\Mc}{\mathcal{M}}

\newcommand{\Oc}{\mathcal{O}}

\newcommand{\Cbb}{\mathbb{C}}
\newcommand{\Ibb}{\mathbb{I}}
\newcommand{\Jbb}{\mathbb{J}}

\newcommand{\sigmab}{\boldsymbol{\sigma}}

\newcommand{\ab}{\mathbf{a}}
\newcommand{\bb}{\mathbf{b}}
\newcommand{\eb}{\mathbf{e}}
\newcommand{\fb}{\mathbf{f}}

\newcommand{\nb}{\mathbf{n}}

\newcommand{\lb}{\mathbf{l}}
\newcommand{\db}{\mathbf{d}}
\renewcommand{\sb}{\mathbf{s}}

\newcommand{\Ab}{\mathbf{A}}
\newcommand{\Bb}{\mathbf{B}}
\newcommand{\Cb}{\mathbf{C}}
\newcommand{\Db}{\mathbf{D}}
\newcommand{\Gb}{\mathbf{G}}
\newcommand{\Lb}{\mathbf{L}}

\newcommand{\Fb}{\mathbf{F}}

\newcommand{\Mb}{\mathbf{M}}

\newcommand{\Eb}{\mathbf{E}}

\newcommand{\Pb}{\mathbf{P}}

\newcommand{\Ib}{\mathbf{I}}

\newcommand{\Tb}{\mathbf{T}}

\newcommand{\tr}{\operatorname{tr}}

\newcommand{\dev}{\operatorname{dev}}
\newcommand{\sym}{\operatorname{sym}}

\title{\bf Machine learning models of plastic flow based on representation theory}
\author{
R.E. Jones\thanks{corresponding author: rjones@sandia.gov}\\
{\it\small Mechanics of Materials Department,}\\
{\it\small Sandia National Laboratories, P.O. Box 969, Livermore, CA 94551, USA} \\
J.A. Templeton\\
{\it\small Thermal/Fluid Science and Engineering Department,}\\
{\it\small Sandia National Laboratories, P.O. Box 969, Livermore, CA 94551, USA} \\
C.M. Sanders \\
{\it\small Thermal/Fluid Science and Engineering Department,}\\
{\it\small Sandia National Laboratories, P.O. Box 969, Livermore, CA 94551, USA} \\
J.T. Ostien\\
{\it\small Mechanics of Materials Department,}\\
{\it\small Sandia National Laboratories, P.O. Box 969, Livermore, CA 94551, USA}
}
\date{}

\begin{document}
\maketitle

\begin{abstract}
We use machine learning (ML) to infer stress and plastic flow rules using data from representative polycrystalline simulations.
In particular, we use so-called {\it deep} (multilayer) neural networks (NN) to represent the two response functions. 
The ML process does not choose appropriate inputs or outputs, rather it is trained on selected inputs and output.
Likewise, its discrimination of features is crucially connected to the chosen input-output map.
Hence, we draw upon classical constitutive modeling to select inputs and enforce well-accepted symmetries and other properties.
With these developments, we enable rapid model building in real-time with experiments, and guide data collection and feature discovery.
\end{abstract}

\section{Introduction}

Our effort to produce viable models of plasticity from trusted data draws upon traditional constitutive modeling theory and newly developed machine learning techniques.

The theory of constitutive function representation has a long history, going back to the beginnings of the Rational Mechanics movement.  
Much of the pioneering work was done by Rivlin, Pipkin, Smith, Spencer, Boehler, and co-workers \cite{spencer1958finite,spencer1958theory,spencer1962isotropic,pipkin1963material,wineman1964material,smith1964integrity,Smith1965,rivlin1969orthogonal,spencer1971part,spencer1987isotropic,boehler1987representations}.
Later, Zheng contributed a notable monograph on the application of representation theory to anisotropy \cite{zheng1994theory}.
Much of these results have been condensed in:
Spencer's monograph \cite{spencer1971part},
Truesdell and Noll's treatise \cite[Sec. 7-13]{truesdell2004non},
Gurtin's text \cite[Sec. 37]{gurtin1982introduction},
and the recent book by Itskov \cite[Ch.4,6,7]{itskov2007tensor}.

The application of machine learning (ML) to engineering dates back to at least the 1980's and covers a wide variety of problems.
For instance,
Adeli and Yeh \cite{adeli1989perceptron} applied ML to the design of steel beams;
Hajela and Berke \cite{hajela1991neurobiological} used a ML model as a surrogate for the exact response of structures to enable fast optimization;
Cheu and Ritchie \cite{cheu1995automated} applied ML to traffic modeling; and
Theocaris and Panagiotopoulos \cite{theocaris1993neural} used it to model fracture behavior and identification.
For further bibliography along these lines, see a review of neural network applications in civil engineering that appeared in 2001  \cite{adeli2001neural}.

Research on applying ML to constitutive modeling dates back to roughly the same time period.
In solid mechanics in particular, 
Ghaboussi \etal \cite{ghaboussi1998autoprogressive} applied a  neural network (NN) to data from experiments of beam deflection.
They created a model which acquired increasing fidelity as experiment progressed via hierarchical learning and adapting new hidden layers.
Furukawa and Yagawa \cite{furukawa1998implicit} constructed an ``implicit'' model of linear viscoplasticity with a NN based on a state space formulation, where the NN provided the driving term for plastic evolution and the elastic response was assumed to be known.
Notably, they expressed a need for variety in the training data.

More recently, a number of studies have appeared comparing NN plasticity models to other models calibrated on experimental data for specific materials.
Lin \etal \cite{lin2008application} built a NN model of the flow stress of low alloy steel based on only experimentally observable quantities.
Bobbili \etal \cite{bobbili2015prediction} constructed a NN model of high strain rate Hopkinson bar tests of 7017 aluminium alloy and compared it to a Johnson-Cook model.
For T24 steel,
Li \etal \cite{li2012comparative} compared a NN model to a modified Zerilli-Armstrong and strain-compensated Arrhenius-type model.
They remarked on the opacity of the NN model and the need for extensive data.
Desu \etal \cite{desu2014support} made flow stress prediction of austenitic 304 stainless steel with support vector machine construct and compared it to a NN model.
Asgharzadeh \etal \cite{asgharzadeh2016study} modelled the flow stress behavior of AA5086 aluminum using a NN with two hidden layers.
(Also, in the realm of fluid mechanics, Ling \etal \cite{ling2015evaluation,ling2016reynolds}, Duraisamy \etal \cite{tracey2015machine,duraisamy2015new}, and Koumoutsakos \etal \cite{milano2002neural} have been particularly active in applying machine learning techniques to model turbulence \cite{wang2017mlturbulence}.)
Unlike traditional models based on physical mechanisms and intuition, these ML models are purely data-driven and phenomenological.
Recently, mathematical analysis has been applied to understanding the training and response structure of NNs, which have traditionally been treated as black boxes. 
The work of Tishby and co-workers \cite{shwartz2017opening} (and Koh and Liang \cite{koh2017understanding}) is particularly illuminating and explores the trade-offs between information compression and prediction accuracy in the training process.

In the wider context of data-driven modeling, a number of recent developments \cite{alharbi2015crystal,kirchdoerfer2016data,smith2016linking,versino2017data,bessa2017framework} are also noteworthy.
Alharbi and Kalidindi \cite{alharbi2015crystal} constructed a database of Fourier transformed microstructural data and used this spectral information to drive evolution of crystal plasticity simulation.
Kirchdoerfer and Ortiz \cite{kirchdoerfer2016data} sought to subvert the traditional empirical model in the data-to-model-to-prediction chain and replace it with a penalization of the prediction response by its distance to closest experimental observation/data point.
This approach of directly using a database is commendable (but lacked data interpolation which appears,  for example, in \cref{shaughnessy2016efficient}).
The optimization was constrained by conservation principles like a Newtonian force balance and was applied to truss and elasticity problems.
The authors explored the technique's robustness to noise and convergence.
Versino \etal \cite{versino2017data} applied a genetic/evolutionary algorithm and a symbolic regression to model Taylor impact test data.
The symbolic regression machine learning technique selects a best model composed of given analytic building-blocks and is especially attractive since the resulting tree structure leads to a physically intepretable model based on the physics embedded in the building-block sub-models.
Lastly, Bessa \etal \cite{bessa2017framework} integrated design of experiments, simulation, and machine learning in materials discovery and design.
It should be noted that the Materials Genome and similar material discovery and selection efforts \cite{jain2013commentary,saal2013materials,raccuglia2016machine} are a  deep and active field of research but this  classification problem has minor bearing on the constitutive modeling task at hand.

In the vein of designing the architecture NN suit to specific tasks, the method we adopt and generalize, the Tensor Basis Neural Network (TBNN) \cite{ling2016machine}, is not simply a feed-forward, deep neural network.
Unlike other NN mechanics models of components of output quantities, \eg stress, TBNN models have built-in invariance properties.
The TBNN formulation shifts the basis for the unknown coefficient functions from the (arbitrary) Cartesian basis of the training data to an objective basis made up of powers of the selected inputs, as representation theory \cite{spencer1971part,truesdell2004non} suggests.
This comes with the cost that the coefficient functions and basis are not linearly independent \ie they must be trained simultaneously.
This representation is akin to the Gaussian Approximation Potential (GAP) with the Smooth Overlap of Atomic Positions (SOAP) basis \cite{bartok2015gaussian} that is gaining popularity in molecular dynamics, in that this machine learning constitutive function uses a spectral basis to preserve rotational and permutational invariance.
It also has goals in common with image transforms that embed invariance properties \cite{khotanzad1990invariant,lowe1999object}.

Motivated by the goal of achieving on-the-fly model construction,  directed sampling/experiments, and discovery of features/trends in large datasets, in this work we show how classical constitutive modeling is needed to obtain viable ML models of constitutive behavior.
In \sref{sec:theory}, we provide the fundamentals of representation and plasticity theories and connect them with our NN formulation of the components of plasticity, namely the stress and flow rules.
In \sref{sec:methods}, we discuss how the data to train the models is obtained, the specifics of the learning algorithm, and the time integration algorithm used to predict the plastic evolution.
One of the data sets is obtained from the elastic-plastic response of an ensemble of oligo-crystalline aggregates, and so the resulting NN model can be considered a form of homogenization.
The results of these developments are discussed \sref{sec:results} and include comparisons of various model architectures and inputs based on cross-validation errors and evaluations of stability and prediction accuracy.
Finally, in \sref{sec:discussion}, we discuss results and innovations, such as the generalized tensor basis architecture, the novel ways of embedding physical constraints in the formulation, and the exploration of data sufficiency, robustness, and stability.

\section{Theory} \label{sec:theory}

In this section we provide a concise overview of representation theory and how we apply it in the context of constitutive modeling by (artificial) neural networks (NNs).
Specifically, we employ a generalization of the Tensor Basis Neural Network (TBNN) \cite{ling2016machine} concept based on an understanding of classical representation theory.
With it we construct models that represent the selected output as a function of inputs with complete generality and compact simplicity.
This construction is distinct from the predominance of component-based NN constructions, for example those mentioned in the Introduction, in that basic symmetries, such as frame invariance are built in to the representation and do not need to be learned.

\subsection{Representation theory}

Representation theorems for functions of tensors have a foundation in group theory \cite{olver2000applications,goodman1998representations,goodman2009symmetry,sattinger2013lie} with the connection being that symmetry is described as functional invariance  under group action.
In mechanics, the relevant invariance under group action are rotations (and translations) of the coordinate system, which is known as {\it material frame indifference}, {\it invariance under super-posed rigid body motions} or simply {\it objectivity}.%
\fnote{Frame indifference is a special case of the  more general principle of covariance with changes of the metric tensor \cite[Sec.3.3]{marsden1994mathematical}. }
This is a fundamental and exact symmetry.
Practical applications of representation theory to mechanics are given in Truesdell and Noll's monograph \cite[Sec. 7-13]{truesdell2004non} and Gurtin's text \cite[Sec. 37]{gurtin1982introduction} and address complete, irreducible representations of general functions of physical vector and tensor arguments.
For example, the scalar function $f(\Ab)$ of a (second order) tensor $\Ab$ is invariant if
\begin{equation} \label{eq:scalar_invariance}
f(\Ab) = f(\Gb \Ab \Gb^T) \ ,
\end{equation}
and a (second order) tensor-valued function $\Mb(\Ab)$ is objective if
\begin{equation} \label{eq:tensor_invariance}
\Gb \Mb(\Ab) \Gb^T = \Mb(\Gb \Ab \Gb^T) \ ,
\end{equation}
for every member $\Gb$ of the orthogonal group.

Underpinning the representations of $f$ and $\Mb$ are a number of theorems.
The {\it spectral theorem} states that any {\it symmetric} second order tensor $\Ab$ has spectral representation :
\begin{equation} \label{eq:spectral}
\Ab = \sum_{i=1}^3 \lambda_i \ab_i \otimes \ab_i \ ,
\end{equation}
composed of  its eigen-values $\{ \lambda_i \}$ and eigen-vectors $\{ \ab_i \}$ where $i=1,3$.
The spectral representation of $\Ab$ makes powers of $\Ab$ take a simple form: $\Ab^n = \sum_i \lambda^n_i \ab_i \otimes \ab_i$ (and in particular $\Ab^0 \equiv \Ib$).
The equally important Cayley-Hamilton theorem states that the tensor $\Ab$ satisfies its characteristic equation :
\begin{equation} \label{eq:cayley}
\Ab^3 
- \underbrace{( \lambda_1 + \lambda_2 + \lambda_3 )}_{J_1 = \tr \Ab} \Ab^2 
+ \underbrace{(\lambda_1 \lambda_2 + \lambda_2 \lambda_3 + \lambda_3 \lambda_1)}_{J_2 = \frac{1}{2} \left( \tr^2 \Ab - \tr \Ab^2 \right)} \Ab 
- \underbrace{(\lambda_1 \lambda_2 \lambda_3)}_{J_3=\det\Ab} \Ib = \mathbf{0} \ ,
\end{equation}
where $\{ J_i \}$ are the {\it principal  (scalar) invariants} of $\Ab$.
The (generalized) Rivlin's identities \cite{rivlin1955further,rivlin1997identities} provide similar relations for multiple tensors and their joint invariants.

Scalars that respect \eref{eq:scalar_invariance}, such as $\{ J_i \}$, are called  {\it scalar invariants} and are formed from (polynomials or, more generally, functions of) the eigenvalues of $\Ab$.
Hence, $f(\Ab)$ reduces to
\begin{equation} \label{eq:scalar_f}
f(\Ab) = f(\Ic)
\end{equation}
where $\Ic$ is a set of scalar invariants of $\Ab$, and hence $f$ is also an invariant.
A set of invariants $\Ic$ is considered {\it irreducible} if each of its elements cannot be represented in terms of others and conveys a sense of completeness and simplicity.%
\fnote{In some sense, a complete set of invariants are coordinates on the manifold induced by symmetry constraints and hence are clearly not unique in their ability to coordinatize the manifold.}
Since the eigenvalues $\{\lambda_i\}$ are costly to compute, typically traces such as $\{ \tr \Ab, \tr \Ab^2, \tr \Ab^3 \} = \{ \sum_i \lambda_i, \sum_i \lambda_i^2, \sum_i \lambda_i^3\}$ are employed as scalar invariants.
Joint invariants of a functional basis for multiple arguments are formed with the help of Pascal's triangle.

For tensor-valued functions such as $\Mb(\Ab)$ in \eref{eq:tensor_invariance}, a power series representation 
\begin{equation}
\Mb(\Ab) = \sum_{i=0}^\infty c_i(\Ic) \Ab^i
\end{equation}
is a good starting point.
The coefficient functions $c_i$ are represented in terms of scalar invariants as in \eref{eq:scalar_f}.
This power series representation can be reduced by application of the Cayley-Hamilton theorem \eqref{eq:cayley}, in the recursive form $ \Ab^{j+3} =   J_1 \Ab^{j+2} - J_2 \Ab^{j+1} + J_3 \Ab^j$. 
The {\it transfer theorem} (as referred to by Gurtin \cite[Sec. 37]{gurtin1982introduction}) states that isotropic functions such as $\Mb(\Ab)$ inherit the eigenvalues of their arguments and implies the fact that these functions are co-linear with their arguments.
Also Wang's lemma ($\Ib,\Ab,\Ab^2$ span the space of all tensors co-linear with $\Ab$) is a consequence of \eref{eq:spectral} and \eref{eq:cayley}, and gives a sense of completeness of the representation:
\begin{equation} \label{eq:representation1}
\Mb(\Ab) = c_0(\Ic) \Ib + c_1(\Ic) \Ab + c_2(\Ic) \Ab^2 \ .
\end{equation}
\eref{eq:representation1} evokes the general representation for a symmetric tensor function of an arbitrary number of arguments in terms of a sum of scalar coefficient functions multiplying the corresponding elements of the tensor basis.
The general methodology for constructing the functional basis to represent scalar functions is given in Rivlin and Ericksen \cite{rivlin1955stress}, and the corresponding methodology to construct tensor bases is developed in Wang \cite{wang1969general,wang1970new}.

Representation theory, like machine learning, does not determine the {\it appropriate} arguments/inputs and output for the constitutive functions.
In mechanics, there is a certain amount of fungibility to both.
For instance, the (spatial) Cauchy stress can easily be transformed into the (referential) first Piola-Kirchhoff stress, and left and right Cauchy-Green stretch have same eigenvalues but different eigen-bases.
Also, any of the Seth-Hill/Doyle-Ericksen strain family \cite{seth1961generalized,hill1968constitutive,doyle1956nonlinear} provide equivalent information on deformation, and any of the objective rates formed from Lie derivatives \cite{johnson1984discussion,szabo1989comparison,haupt1989application,haupt1996stress} provide equivalent measures of rate of deformation; however, some choices of arguments and output lead to greater simplicity than others.

Lastly, it is important to note that isotropic functions are not restricted to isotropic responses.
The addition of a structure tensor characterizing the material symmetry to the arguments allows isotropic function theory to be applied so that the joint invariants encode anisotropies \cite{smith1957stress,smith1957anisotropic,spencer1982formulation, zhang1990structural, svendsen1994representation, zheng1994theory}.

\subsection{Plasticity models} \label{sec:plasticity}
Briefly, plasticity is an inelastic, history-dependent process due to dislocation motion or other dissipative phenomena.
We assume the usual multiplicative decomposition \cite{lee1969elastic,lubarda2004constitutive} of the total deformation gradient $\Fb$ into elastic (reversible) $\Fb_e$ and plastic (irreversible) $\Fb_p$ components
\begin{equation} \label{eq:FeFp} 
\Fb=\Fb_e \Fb_p  \ .
\end{equation}
As a consequence, the velocity gradient in the current configuration, $\lb \equiv \dot{\Fb} \Fb^{-1}$, can be additively decomposed into elastic and plastic components :
\begin{equation}
  \lb=\dot{\Fb}_e\Fb_e^{-1}+\Fb_e\underbrace{\dot{\Fb}_p\Fb_p^{-1}}_{{\Lb}_p}\Fb_e^{-1} \ ,
\end{equation}
refer to \cite[Sec. 8.2]{Lubliner2008}.
The assumption that $\Fb_p$ is pure stretch (no rotation) reduces $\Lb_p$ to $\Db_p = \sym \Lb_p$.
The elastic deformation determines the stress, for instance the Cauchy stress $\Tb$:
\begin{equation} \label{eq:stress}
\Tb = \hat\Tb(\Fb_e) = \Tb(\eb_e) \ ,  
\end{equation}
and the evolution of the  plastic state is determined by a {\it flow rule}, \eg :
\begin{equation} \label{eq:dotFp}
\dot{\Fb}_p  = \Db_p \Fb_p \ \ \text{where} \ \
\Db_p = \hat\Db_p(\Fb_p,\Tb) = \Db_p(\bb_p,\sigmab) \ ,
\end{equation}
where $\Fb_p$ quantifies the plastic state and $\Tb$ the driving stress. 
Invariance allows the reduction of the argument of $\Tb$ to, for example, the objective, elastic Almansi strain $\eb_e = \frac{1}{2} \left( \Ib - \bb_e^{-1} \right)$ based on the left Cauchy-Green/Finger stretch tensor $\bb_e = \Fb_e \Fb_e^T$. 
Similarly, the state variable in the flow rule can be reduced by applying invariance, for example, $\bb_p = \Fb_p \Fb_p^T$.
The driving stress can be attributed to the deviatoric part of the pull-back of the Cauchy stress $\Tb$: 
$\sigmab = \dev \left[ \Fb_e^{-1} \Tb \Fb_e^{-T} \right]$ which is also invariant and also coexists in the intermediate configuration with $\Db_p$.
Furthermore, a deviatoric tensor basis element, such at $\sigmab$, generates an isochoric flow which respects plastic incompressibility $\det \Fb_p \equiv 1$.
Other choices of the inputs and outputs of the stress and flow functions are discussed in Results section.
Typically both the stress and flow are derived potentials to ensure elastic energy conservation for the stress and associative flow for the flow rule; however, in this work we  to allow for a more general flow and non-differentiable NN model.
(Experiments typically cannot measure potentials directly).%
\fnote{Also worth mentioning are the complex requirements for elastic stability, see Ref. \cite[Sec. 5]{marsden1994mathematical}, that we do not attempt to embed in the formulation mainly because they require a potential.}

A few basic properties are built into traditional empirical models that need to be learned in typical NN models.
First, zero strain, $\eb_e = \mathbf{0}$, implies zero stress :
\begin{equation} \label{eq:zero_stress}
\Tb(\mathbf{0}) = \mathbf{0} \ ,
\end{equation}
and, likewise, zero driving stress should result in zero plastic flow :
\begin{equation} \label{eq:zero_flow}
\Db_p(\Fb_p,\mathbf{0}) = \mathbf{0} \ .
\end{equation}
Also there is a dissipation requirement for the plastic flow.
Generally speaking, the Coleman-Noll \cite{coleman1963thermodynamics} argument, together with the first and second law of thermodynamics,  applied to a free energy in terms of the elastic deformation and a plastic history variable results in: (a) the stress being conjugate to the elastic strain rate, and (b) the internal, plastic state variable, when it evolves, reduces the free energy via $ \Mb\cdot\Lb_p \ge 0 $ where $\Mb$ is the Mandel stress 
\begin{equation} \label{eq:mandel}
\Mb = \det(\Fb) \, \left[ \Fb_e^T \Fb_e \right]  \left[ \Fb^{-1} \Tb \Fb^{-T} \right] 
\end{equation}
This reduces to 
\begin{equation} \label{eq:plastic_dissipation}
 \Tb \cdot \db_p \ge 0  \ ,
\end{equation}
refer to Ref. \cite[Sec. 8.2]{Lubliner2008}.
Also, given the physics of dislocation motion, it is commonly assumed that the plastic deformation is  incompressible,
$\det \Fb_p = 1$, which implies the flow is deviatoric  
\begin{equation} \label{eq:plastic_incompressibility}
\tr \Db_p = 0
\end{equation}
For more details see the texts \crefs{Lubliner2008,Simo1998,Gurtin2010}.

\subsection{Application to neural network constitutive modeling} \label{sec:tbnn}
We generalize the Tensor Basis Neural Network (TBNN) formulation \cite{ling2016machine} to build  NN representations for the stress relation, \eref{eq:stress}, and the plastic flow rule, \eref{eq:dotFp},  that embed a number of symmetries and constraints.
Both $\Tb$ and $\Db_p$ are required to be isotropic functions of their arguments by invariance.
As discussed, classical representation theorems give the general form
\begin{equation} \label{eq:representation}
\fb(\Ac) = \sum_i f_i(\Ic) \, \Bb_i \ ,
\end{equation}
where $\Ac\equiv\{\Ab_1,\Ab_2,\ldots\}$ are the pre-supposed dependencies/arguments of function $\fb$,  $\Ic \equiv \{I_j\}$ is an (irreducible) set of scalar invariants of $\Ac$, and $\Bc = \{ \Bb_j \}$ is the corresponding tensor basis.
In \eref{eq:representation}, only the scalar coefficient functions are $\{ f_i \}$ are unknown once the inputs have been selected and hence they are represented with a dense NN using the selected scalar invariants $\Ic$ as inputs embedded in the overall TBNN structure.
In the TBNN framework, the sum the NN functions $\{ f_i(\Ic) \}$ and the corresponding tensor basis elements $\{ \Bb_i \}$ in \eref{eq:representation} is accomplished by a so-called {\it merge layer}, and the functions $\{ f_i \}$  are trained simultaneously (refer to \fref{fig:tbnn} and more details will be given in \sref{sec:training}).
This formulation is in contrast to the standard, component-wise NN formulation:
\begin{equation} \label{eq:cnn_representation}
\fb(\Ac) = \sum_{i,j} f_{ij}(\left[\Ab_1\right]_{ij}, \left[\Ab_2\right]_{ij}, \ldots) \, \eb_i \otimes \eb_j \ ,
\end{equation}
which is based on components of both the inputs $\{\Ab_1,\Ab_2,\ldots\}$ and the output $\fb$.

For the stress, we assume a single symmetric tensor input selected from the Seth-Hill/Doyle-Ericksen elastic strain family, in particular $\eb_e$, is sufficient, so that representation \eref{eq:representation1}:
\begin{equation} 
\Tb = \stress_0(\Ic) \Ib + \stress_1(\Ic) \eb_e + \stress_2(\Ic) \eb_e^2 \ ,
\end{equation}
is appropriate.
Despite this formulation being based on strain, versus stretch, it does not embed the zero stress property, \eref{eq:zero_stress}, and, hence, $\stress_0(\Ic)$ will need to learn that zero strain implies zero stress.
Since we prefer to impose, rather than learn, physical constraints such as \eref{eq:zero_stress} since this reduces the necessary training data \cite{ling2016machine} and the exact satisfaction leads to conservation and other properties necessary for stability, \etc 
Exact satisfaction of \eref{eq:zero_stress} can accomplished a few different ways:
(a) shifting the basis with the Cayley-Hamilton theorem \eqref{eq:cayley}
\begin{equation}
\Tb = \stress_1 \eb_e + \stress_2 \eb_e^2 + \stress_3 \eb_e^3 \ ,
\end{equation}
refactoring (b) some 
$
\Tb = \left( I_2 \stress'_0 \right) \Ib + \stress'_1 \eb_e + \stress'_2 \eb_e^2
$, 
or (c) all 
$
\Tb = I_2 \left( \stress_0'' \Ib + \stress_1'' \eb_e + \stress_2'' \eb_e^2 \right)
$ 
of the coefficient functions $\{ \stress_i \}$ with $I_2 = \tr \eb_e^2$.
In general, any of these representations can be expressed on the spectral basis
\begin{equation} \label{eq:spectral_components}
\Tb =  \sum_i \sum_{j=1}^3 \sigma_i \lambda^i_j \ab_j \otimes \ab_j
    =  \sum_{j=1}^3 \left( \sum_i \sigma_i \lambda^i_j \right) \ab_j \otimes \ab_j
\end{equation}
so there is a (weak) equivalence between coefficient functions of the various representations.
Here, $\eb_e = \sum_i \lambda_i \ab_i \otimes \ab_i$.

As mentioned, we assume that the inputs to the flow rule are (a) a history variable $\bb_p$, and (b) driving stress $\sigmab$.
A general function representation from classical theory for an isotropic function of two (symmetric) tensor arguments requires ten invariants \cite{rivlin1955further} (see also \cite[Ch.3, Eq. 9 and 11]{boehler1987representations}):
\begin{equation}
\Ic \equiv \{ I_i \}
= \{
\tr \bb_p, \tr \bb_p^2, \tr \bb_p^3, 
\tr \sigmab, \tr \sigmab^2, \tr \sigmab^3,
\tr \bb_p \sigmab, \tr \bb_p^2 \sigmab, \tr \bb_p \sigmab^2, \tr \bb_p^2 \sigmab^2
\}
\end{equation}
and eight tensor generators/basis elements
\begin{equation}
\Bc
\equiv \{ \Bb_i \} 
= \{
\Ib, \bb_p, \bb_p^2, 
     \sigmab, \sigmab^2, 
\sym \bb_p \sigmab,   
\sym \bb_p^2 \sigmab, 
\sym \bb_p \sigmab^2  
\} \ ,
\end{equation}
where $\sym \Ab \equiv \frac{1}{2}(\Ab + \Ab^T)$.
To satisfy the zero flow condition, \eref{eq:zero_flow}, we can shift basis for the second, stress argument and eliminate all basis elements solely dependent on the first, plastic state argument:
\begin{equation}
\Bc
= \{
\sigmab, \sigmab^2,  \sigmab^3,
\sym \bb_p \sigmab,   
\sym \bb_p^2 \sigmab, 
\sym \bb_p \sigmab^2  
\} \ .
\end{equation}
Plastic incompressibility, in the form of deviatoric plastic flow,  \eref{eq:plastic_incompressibility},  can imposed by applying the linear operator $\dev$, $\dev \Ab = \Ab - \frac{1}{3}\tr(\Ab)\Ib$,
\begin{align}
\Db_p &= 
f_{01} \dev \sigmab + f_{11} \sym \dev \bb_p \sigmab+ f_{02} \dev \sigmab^2 \nonumber\\
&+  
f_{21} \dev \sym \bb_p^2 \sigmab + f_{12} \dev \sym \bb_p \sigmab^2 
\nonumber
\end{align}
Dissipation of plastic flow can be strictly imposed by requiring that the flow be directly opposed to the stress in \eref{eq:plastic_dissipation} which implies:
\begin{equation}
\Db_p = f_1 \,  \sigmab + f_3 \, \sigmab^3 \ , 
\end{equation}
and $f_1(\Ic) > 0$ and  $f_3(\Ic) > 0$.
In this study we will rely on the learning process to ensure the positivity of the coefficient functions $f_1$ and $f_3$ but this could be accomplished exactly with the Macauley bracket (ramp function) applied to $f_1$ and $f_3$, for example.

\section{Methods} \label{sec:methods}

We train the NN models of plasticity with data from two traditional plasticity models.
In this section we give details of (a) the traditional models, (b) the training of the NNs, and (c) numerical integration of the TBNN plasticity model.

\subsection{Plasticity models} \label{sec:plasticity_models}
In an exploration of the fundamental properties of NNs applied to plasticity,   we seek to represent responses of two models: 
(a) a poly-crystalline representative volume element (RVE) with grain-wise crystal plasticity (CP) response  (an {\it unknown} closed form model since the poly-crystalline aspect of the CP model obscures its closed form), and (b) a simple visco-plasticity (VP) material point  (a {\it known} closed form model).
Both are finite deformation models so that invariance and finite rotation are important; and both are visco-plastic in the sense of lacking a well-defined yield surface and strictly dissipative character.

Briefly, crystal plasticity (CP) is a well-known meso-scale model of single crystal deformation. 
Here we use crystal plasticity to prescribe the response of individual crystals in a perfectly bonded polycrystalline aggregate.
The theoretical development of CP is described in \crefs{taylor1934mechanism,kroner1961plastic,bishop1951xlvi,bishop1951cxxviii,mandel1965generalisation} and the computational aspects in reviews \cite{dawson2000computational,roters2010overview}.

Specifically, for the crystal elasticity, we employ a St. Venant stress rule formulated with the second Piola-Kirchhoff stress mapped to the current configuration
\begin{equation} \label{eq:unknown_stress}
\Tb = \frac{1}{\det \Fb} \Fb \left(  \Cbb \Eb_e \right) \Fb^T
\end{equation}
where the elastic modulus tensor $\Cbb = C_{11} \Jbb  + C_{12} ( \Ibb -\Jbb ) + C_{44} ( \Ib \otimes \Ib -\Jbb)$ has cubic crystal symmetries with $C_{11}, C_{12}, C_{44}$ = 204.6, 137.7, 126.2 GPa, and $\Eb_e = \frac{1}{2}\left(\Fb_e^T \Fb_e - \Ib \right)$ is the elastic Lagrange strain.
Here 
$\left[ \Jbb \right]_{ijkl} = \delta_{ij} \delta_{kl} \delta_{ik} \delta_{jl}$,
$\left[\Ibb\right]_{ijkl} = \frac{1}{2} \left(  \delta_{ik} \delta_{jl} + \delta_{il} \delta_{jk} \right)$
and $\delta_{ij}$ is the Kronecker delta.
Plastic flow can occur on any of 12 face-centered cubic (FCC) slip planes.
Each crystallographic slip system, indexed by $\alpha$, is characterized by Schmid dyads $ \Pb_{\alpha}=\sb_{\alpha}\otimes\nb_{\alpha} $ composed of the allowed slip direction, $\sb_{\alpha}$, and the normal to the slip plane, $\nb_{\alpha}$.
Given the set $\{ \Pb_{\alpha} \}$, the plastic velocity gradient is constructed via:
\begin{equation} \label{eq:unknown_flow}
  {\Lb}_p=\sum_{\alpha}\dot{\gamma}_{\alpha}\Pb_{\alpha} \ ,
\end{equation}
which is inherently volume preserving in the (incompatible) intermediate/lattice configuration.
Finally, the slip rate $\dot{\gamma}_{\alpha}$ is related to the applied stress through the resolved shear (Mandel) stress $ \tau_{\alpha}=\Mb\cdot\Pb_{\alpha}, $ for that slip system.
We employ a common power-law form for the slip rate relation 
\begin{equation}
  \dot{\gamma}_{\alpha}=\dot{\gamma}_{\alpha 0}\left|\frac{\tau_{\alpha}}{g_{\alpha}}\right|^{1/m}\tau_{\alpha} \ ,
\end{equation}
where $\dot{\gamma}_{\alpha 0}$ = 122.0 (MPa-s)$^{-1}$ is a reference strain rate, $m = 20 $ is a rate sensitivity exponent, and $g_{\alpha}$ = 355.0 MPa is a hardness value.
These parameters are representative of steel.

With this model in Albany \cite{albany}, we simulate the polycrystalline response using a uniform mesh 20 $\times$ 20 $\times$ 20 with the texture assigned element-wise (via Dream3d \cite{dream3d}) and strict compatibility enforced at the voxelated grain boundaries. 
Ten realizations with 15, 15, 17, 18, 18, 19, 19, 20, 20, 21, 22 grains were sampled from an average grain size ensemble and each grain was assigned a random orientation. 
Minimal boundary conditions to apply the various loading modes ( tension, shear, \etc) were employed on the faces and edges of the cubical representative volumes.
Also, we limit samples to a single, constant strain rate 1.0 1/s.

The simple visco-plastic (VP) model consists of a St. Venant stress rule in the current configuration with Almansi strain:
\begin{equation} \label{eq:known_stress}
\Tb = \Cbb \eb_e \ ,
\end{equation}
where
$\Cbb = \lambda \Ib \otimes \Ib + 2 \mu \Ibb$ isotropic parameters $\lambda = \frac{E \nu}{(1+\nu)(1-2\nu)}$ and $\mu = \frac{E}{2(1+\nu)}$ with Young's modulus $E$ = 200 GPa and Poisson's ratio $\nu$ = 0.3, together with 
a simple (associative) power law for the flow rule:
\begin{equation} \label{eq:known_flow}
\Db_p = c \| \sb \|^p \sb \ ,
\end{equation}
where $c$ = 0.001 MPa$^{-1-p}$-s$^{-1}$ and $p$ = 0.1  are material constants.

\subsection{Neural network representation and machine learning algorithm} \label{sec:training}

A typical NN, such as the representation of \eref{eq:cnn_representation}, is a two-dimensional feed-forward, directed network consisting of an input layer, output layer and $L$ intervening hidden layers where neighboring layers are densely connected.
Each layer $\ell_i$ consists of $N$ nodes $(ij)$.
The vector of outputs, $\ys_i$, of the nodes $(ij), j \in (1,N)$ of layer $i$ is the weighted sum of the outputs of the previous layer $\ell_{i-1}$ offset by a threshold and passed through a ramp-like or step-like \emph{activation} function $a(x)$:
\begin{equation}
\xs_{i} = a(\ys_{i}) \ \ \text{with} \ \ \ys_{i} = \Ws_{i} \xs_{i-1} + \bs_{i} \ ,
\end{equation}
where $\Ws_i$ is the weight matrix for (hidden) layer $\ell_i$ of the state/output of nodes of the previous layer $\xs_{i-1}$ and $\bs_i$ is the corresponding threshold vector.
In our application the input layer consists of the $N_\Ic$ invariants $\Ic$ and the $N_\Bc$ elements of the tensor basis $\Bc$.
The elements of $\Ic$ form the arguments of the coefficient functions, each having a $L\times N$ neural network representation, while the elements of $\Bc$ pass through the overall network until they are combined with the coefficient functions according to \eref{eq:representation} to form the output via a {\it merge} layer that does the summation.
After exploring the C0 step- and ramp-like {\it rectifying} activation functions commonly used, we employ the ramp-like (C1 continuous) Exponential Linear Unit (ELU) \cite{clevert2015fast} activation function:
\begin{equation}
a(x) = 
\begin{cases}
\exp(x) -1 & \text{if} \ \ x<0 \\
     x    & \text{else} \\
\end{cases}
\end{equation}
to promote smoothness of the response and limit the depth of the network necessary to represent the response relative that necessary with saturating step-like functions.

\begin{figure}
\centering
{\includegraphics[width=\figwidth]{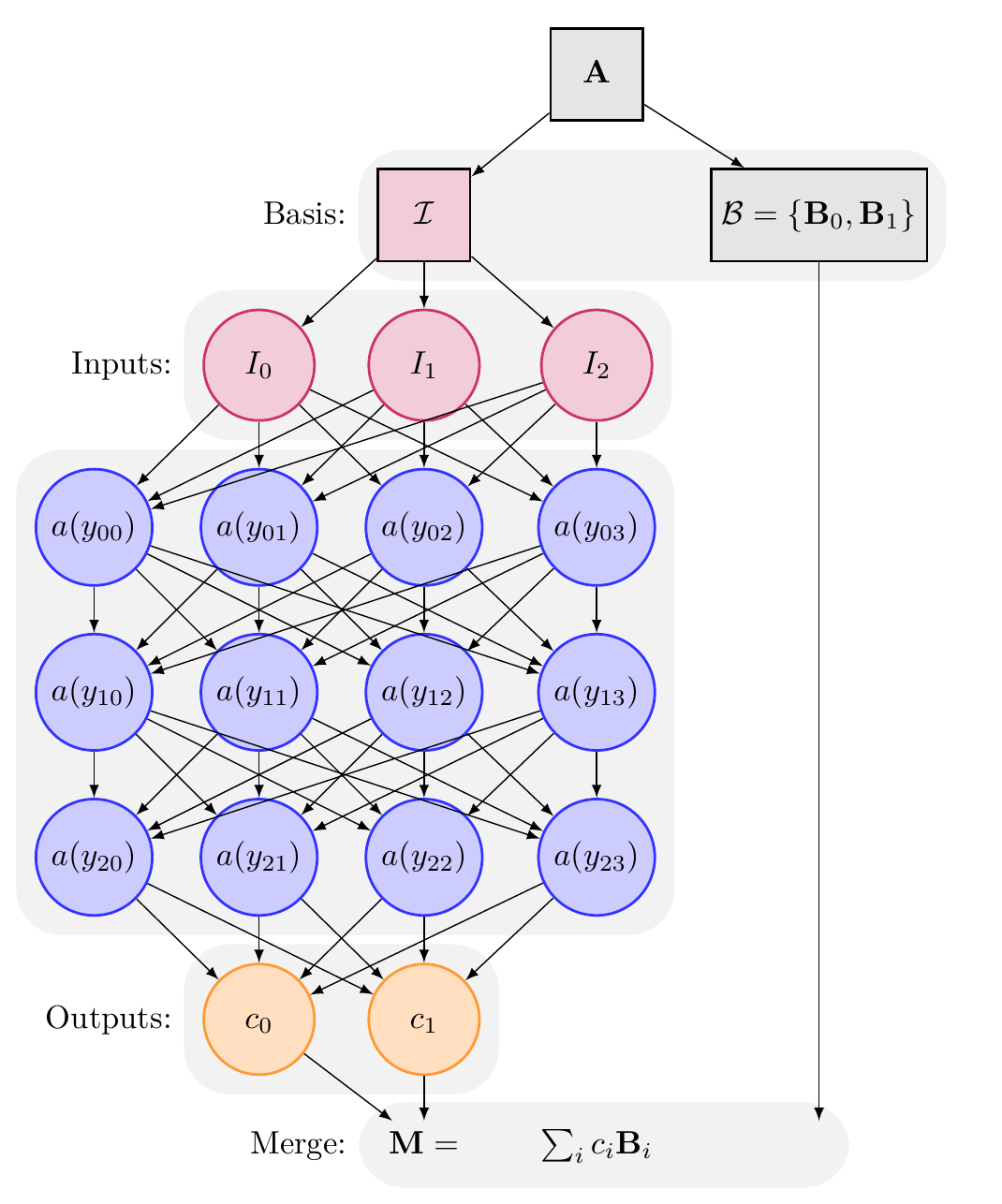}}
\caption{TBNN structure for $\Mb(\Ab) = \sum_i c_i(\Ic) \Bb_i$ with 3 invariants $\Ic = \{I_0,I_1,I_2\}$, a $3\times4$ NN, 2 coefficient functions $\{c_0(\Ic),c_1(\Ic)\}$, and 2 tensor basis elements $\Bc=\{\Bb_0,\Bb_1\}$. 
The scaling operations described in \sref{sec:training} are omitted for clarity.
The linear transformation $\ys_i = \Ws_i \xs_{i-1} + \bs_i$ of the outputs $\xs_{i-1}$  of layer $i-1$ to the inputs $\ys_i$ of layer $i$ is denoted by the arrows connecting the nodes of layer $i-1$ to those of layer $i$.
The nonlinearity of the activation functions $a(\ys_i)$ is represented by $a(y_{ij})$ where $y_{ij}$ are the components of $\ys_i$.
}
\label{fig:tbnn}
\end{figure}

Training the network weights $\Ws_i$ and thresholds $\bs_i$ is accomplished via  the standard back-propagation of errors \cite{werbos1974beyond,rumelhart1986learning} which, in turn, drives a (stochastic) gradient-based descent (SGD) optimization scheme to minimize the so-called {\it loss}/error, $E$.
We employ the usual root mean square error (RMSE) 
\begin{equation} \label{eq:RMSE}
E = \frac{1}{2 N_D} \sum_{ (\xs_k, \ds_k) \in D}  \left\|  \ys(\xs_k) - \ds_k \right\|^2 \ ,
\end{equation}
where $D$ is the set of training data composed of inputs $\xs_k = \{\Ic_k,\Bc_k\}$ and corresponding output $\ds_k$.
The gradient algorithm relies on: (a) the change in $E$ with respect to each weight $\Ws_i$
\begin{equation} \label{eq:derror_dW}
\frac{\partial E}{\partial \Ws_i}  = 
\underbrace{\frac{\partial E}{\partial \xs_i}
\frac{\partial \xs_i}{\partial \ys_i }}_{\Delta_i}
\frac{\partial \ys_i}{\partial \Ws_i}  = 
\xs_{i-1} \otimes \Delta_i
\end{equation}
and (b) each threshold $\bs_i$
\begin{equation} \label{eq:derror_db}
\frac{\partial E}{\partial \bs_i}  = 
\underbrace{\frac{\partial E}{\partial \xs_i}
\frac{\partial \xs_i}{\partial \ys_i }}_{\Delta_i}
\frac{\partial \ys_i}{\partial \bs_i}  = 
\Delta_i \ ,
\end{equation}
where 
\begin{equation} \label{eq:delta_l}
\Delta_i = \left(  \Ws_{i+1}^T \Delta_{i+1} \right) \odot a'(\ys_i)  
\ \ \text{for}  \ \ i \neq L
\ \ \text{with} \ \
\Delta_L = \sum_{(\xs_k,\ds_k) \in D}  \left(  \ys(\xs_k) - \ds_k \right) \odot a'(\ys_L)
\end{equation}
Here $a'$ is the derivative of weight function,
$[\as \otimes \bs ]_{ij}  = a_i b_j $ is the tensor product, and
$[\as \odot \bs ]_i  = a_i b_i $ element-wise Hadamard-Schur product.
The recursion seen in \eref{eq:delta_l} gives back-propagation its name.
The gradient defined by these expressions is evaluated with random sampling of subset of training data $D$ called {\it minibatches}.
Also, search for a minimum along this direction is governed by a step size called the {\it learning rate} in the ML community.
These standard constructions are trivially generalized to the TBNN structure since the inputs $\Bc$ are not directly related to $\Ws_i$ nor $\bs_i$, and are merely scaled by the coefficient functions to form the output $\ys$, refer to \fref{fig:tbnn}.
For more details of the SGD algorithm, see Ref. \cite[Ch.2]{nielsen2015neural}.

To begin the training, the unknown weights, $\{\Ws_i\}$, and thresholds, $\{\bs_i\}$, are initialized with normally distributed random values to break the degeneracy of the network and enable local optimization.
Since multiple local minima for training are known to exist, choosing an ensemble of initial weights which are then optimized improves the chances of finding a global minimum and the distribution of the solutions indicates the robustness of the training.
Also, the full set of input data is divided into a training set $D$, used to generate the errors for the back-propagation algorithm; a test set $T$, for assessing convergence of the descent algorithm; and a third set $V$ for cross-validation, to estimate the predictive capability of the trained network.
Ensuring that the errors based on $T$ are comparable to those on $V$  reduces the likelihood over-fitting data with a larger than necessary NN.
We chose to divide the available data in a $T:D:V =$ 20:72:8 ratio.
In addition, we sample individual stress-strain curves produced by the CP and VP simulators so as to maintain approximate uniform density of data based on curve  arc-length (\vs based on strain) to capture high-gradient (elastic) and transition (yield) regimes.
Also, it should be noted that we allow ourselves to train on inputs derived from the plastic deformation gradient, $\Fb_p$, despite the fact that this quantity is difficult to observe directly in experiments.
A critical part of the training algorithm is normalizing the data so that the NN maps $\Oc(1)$ inputs to $\Oc(1)$ outputs since having $\Ws_i, \bs_i\sim \Oc(1)$ will achieve better SGD convergence.
We also shift and scale the scalar invariants $\Ic$ so that they have a mean zero, variance one distribution.
We normalize the other set of inputs, the tensor basis $\Bc$, using the maximum Frobenius norm  of the basis generators, \eg $\bb_p$ and $\sigmab$, over the training set $D$.
During training, the output tensors are normalized similarly based on their maximum norms over $D$, so that 
\begin{equation} \label{eq:scaling}
\fb = \sum_i \underbrace{ \frac{1}{s_\fb}  \bar{f}_i (\bar{\Ic}) s_{\Bb_i}}_{f_i(\Ic)} \Bb_i \ ,
\end{equation}
where $s_\fb$ is the scaling of output $\fb$;
$s_{\Bb_i}$ is the scaling of basis element $\Bb_i$ based on the powers of principal generator (\eg if $\Bb_i = \bb^a \sb^b$ then $s_{\Bb_i} = s_\bb^a s_\sb^b$ where $s_\bb$ is the scaling of $\bb$); and $\bar{\Ic} = s_{I_i} I_i$ is the set of scaled and shifted invariants.
These scales have the added benefit of coarsely encoding the range of training data so the extrapolation during prediction can be detected.

Convergence is assessed by averaging the error with respect to $T$ over previous iterations of the SDG (in this work we average over the last 4-10 iterations) and terminating when this average converges, but not before performing a minimum number of iterations (1000 in this work).
More discussion of the training approach can be found in \cite{ling2016machine}, although in that work the learning rate was held fixed rather than decaying as the training proceeds, as in this study.

\subsection{Integration algorithm} \label{sec:integration}

We need a time-integration scheme to solve the differential-algebraic system \eref{eq:stress} and \eref{eq:dotFp}.
We assume it is deformation driven so that $\Fb=\Fb(t)$ is data.
To form a numerical integrator, we rely on the well-known exponential map  
\begin{equation} \label{eq:exp_map}
\Fb_{n+\alpha} = \exp\left( \alpha \, \Delta t \left[ \Db_p \right]_n \right) \Fb_{n} 
\end{equation}
which is an explicit/approximate solution to \eref{eq:dotFp}.
In \tref{tab:time_integration} we outline an adaptive scheme based on a midpoint rate at $t_{n+\alpha}$ and interpolation of the deformation gradient:
\begin{equation}
\log \Fb_{n+\alpha} = \log \Fb_n + \alpha \log \Delta \Fb 
 = (1-\alpha) \log \Fb_{n+1} + \alpha \log \Fb_n 
\end{equation}
with $\Delta \Fb = \Fb_{n+1} \Fb_n^{-1}$ so $\Fb_{n+\alpha} = \exp( \alpha \log \Delta \Fb) \, \Fb_n$.
Since we do not rely on the NN models of stress \eref{eq:stress} and flow \eqref{eq:dotFp} being directly differentiable,%
\footnote{This relaxation could be improved by using the derivatives already computed by the backpropagation algorithm in a Newton solver with a trust region based on the bounds of the training data.}
we use a simple relaxation scheme to enforce consistency:
\begin{equation} \label{eq:consistency}
\left[ \Fb_p \right]_{n+1} = \exp\left(
\Delta t \,
\Db_p \left\lgroup
\left[ \Fb_p \Fb_p^T \right]_n, 
\dev \Tb\left(\frac{1}{2} \left( \Ib - \Fb^{-T} \left[ \Fb^T_p \Fb_p \right]_n^{-1} \Fb^{-1} \right) \right)
\right\rgroup
\right) \left[ \Fb_p \right]_{n}
\end{equation}
for $\left[\Fb_p \right]_{n+1}$ given $\left[\Fb_p \right]_n$ and $\Fb\equiv\Fb_{n+1}=\Fb(t_{n+1})$.
Here we have simply substituted stress and flow rules into \eref{eq:exp_map} with the particular arguments $\Tb(\eb_e)$ and  $\Db_p(\bb_p,\sb)$.
If any step has an increase in error formed from the residual of \eref{eq:consistency} the step size is cut; and, conversely, when a sub-step converges, the remainder of the interval is attempted.

\begin{table}
\centering
\fbox{\parbox{0.95\textwidth}{
\footnotesize
For step $n+1$
\begin{itemize}
\item Initialize $\Fb = \Fb_{n}$  and $\Delta \Fb = \Fb_{n+1} \Fb_{n}^{-1}$
\item Sub-step: while $\alpha < 1$
\item Try $\alpha = 1$, $\Fb_{n+\alpha} = \exp\left( \alpha \log\left( \Delta \Fb \right) \right) \, \Fb_n$
\begin{itemize}
\item Relaxation: loop over $k$, initialize $\left[\Fb_p \right]^*_{k=0} = \Fb_p $:
\begin{enumerate}
\item $\bb_p^* = \left[  \Fb \Fb^T \right]_k^*$ and $\bb_e^* = \Fb \left[ \left(\Fb_p^T \Fb_p\right)^{-1} \right]_k^* \Fb^T$
\item $\Tb^* = \Tb(\bb_e^*)$ and
      $\sb^* = \frac{1}{\det{\Fb}} \dev \Tb^*$
\item $\left[ \Db_p \right]_k^* = \fb(\bb_p^*,\sb^*)$ 
\item $\left[ \Fb_p \right]_{n+\alpha} = \exp\left( \alpha \Delta t \, \Db_p^*   \right) \, \left[ \Fb_p \right]_n$
\item if $\left\| \left[\Db_p \right]^*_{k} - \left[\Db_p \right]^*_{k-1} \right\| < \epsilon \left\| \left[ \Db_p \right]_{k-1} \right\|$ then exit, converged \\
else if $\left\| \left[\Db_p \right]^*_{k} - \left[\Db_p \right]^*_{k-1} \right\| >  \left\| \left[\Db_p \right]^*_{k-1} - \left[\Db_p \right]^*_{k-2} \right\|$ then diverging, cut step $\alpha = 1/2 \alpha$ \\
else $\alpha \pluseq \Delta \alpha$ 
\end{enumerate}
\item Update $\Tb_{n+1} = \Tb^*$ and $\left[\Fb_p\right]_{n+1} = \left[\Fb_p \right]^*$
\end{itemize}
\end{itemize}
}}
\caption{Time integration algorithm with adaptive time-stepping.
} \label{tab:time_integration}
\end{table}

\section{Results} \label{sec:results}

In this section we cover our investigations of: (a) optimal network size, inputs, and representation basis; (b) influence of training data on error and stability; and (c) the robustness and accuracy of the model predictions.
As mentioned, we employ data from an unknown-form CP model (\eref{eq:unknown_stress} and \eref{eq:unknown_flow}) and known-form VP model (\eref{eq:known_stress} and \eref{eq:known_flow}).
Training with the data from the CP model illustrates the NN model's ability to represent and homogenize the response of a complex system and the VP model is particularly useful for exploring NN representations since we know the true response and generating samples is computationally inexpensive.

\subsection{Constructing and training the neural networks}

We begin our numerical investigations with:
(a) a survey of the possible representations for the models of stress and flow,
(b) optimizing the structure and meta-parameters of the NN representations.
To assess improvements in performance  we used the traditional metric for evaluating NN performance, {\it cross-validation} error, where the training dataset $D$ replaced by the validation dataset $V$ in evaluating the RMSE formula, \eref{eq:RMSE}.

\begin{figure}[h!]
\centering
\subfloat[][tension]
{\includegraphics[width=\figwidthtwo]{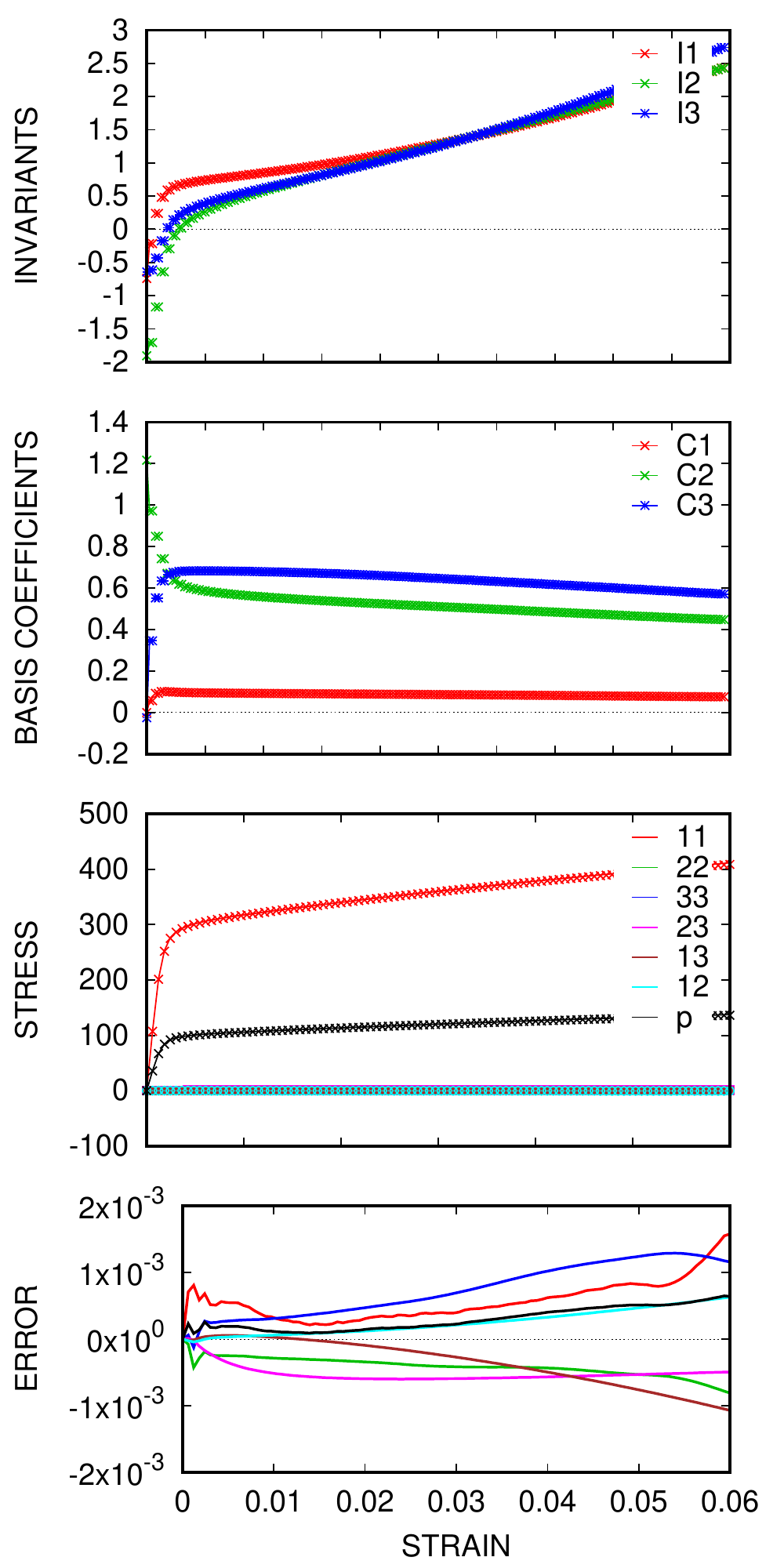}}
\subfloat[][shear]
{\includegraphics[width=\figwidthtwo]{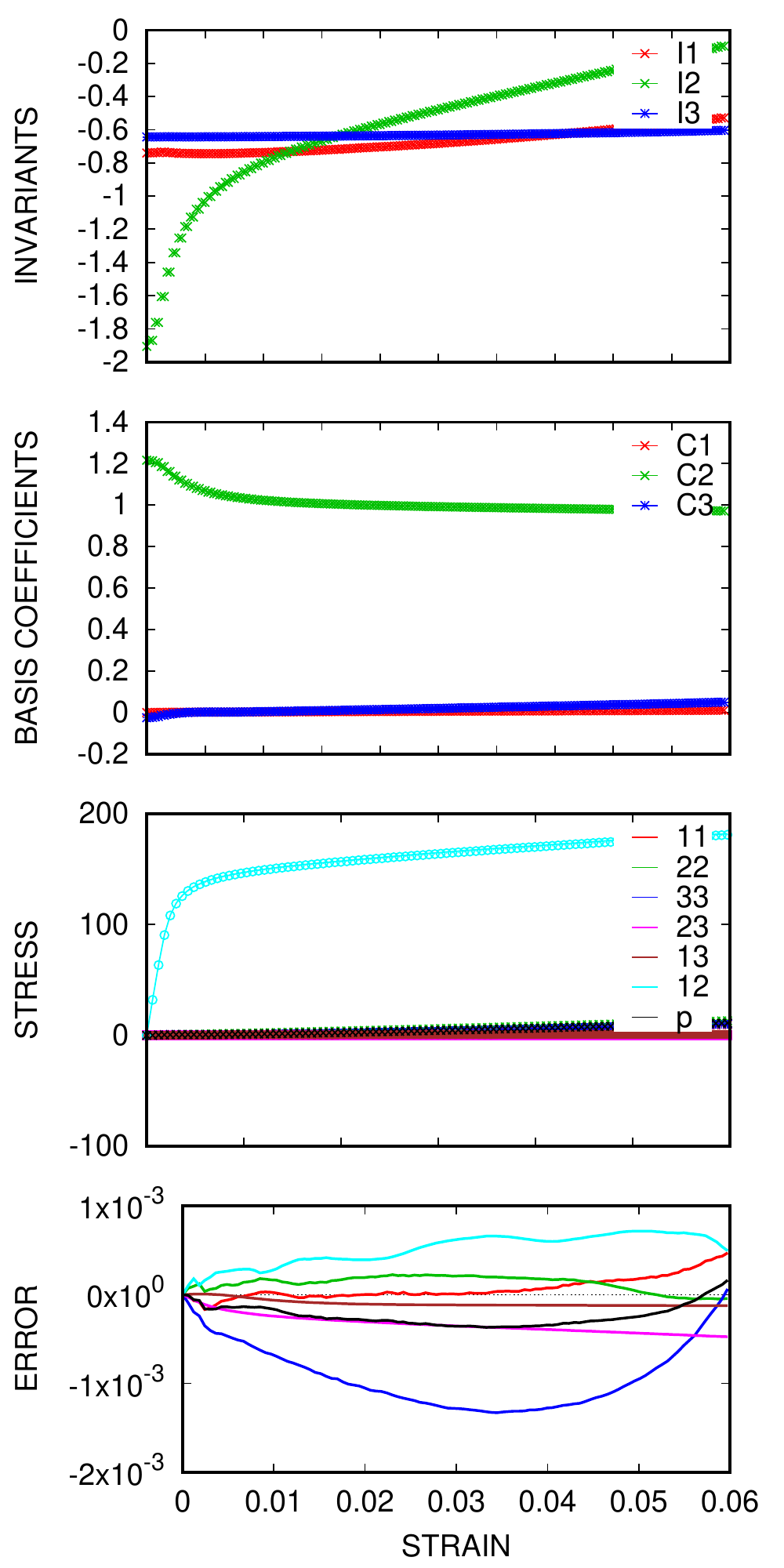}}
\caption{ 
Stress training data: (a) tension and (b) shear stress evolution with strain for CP model.
Top panels: scaled input invariants.
Second panels: scaled trained tensor basis coefficient functions.
Third panels: stress $\Tb$ response (lines: model, points: data).
Bottom panel: error as function of strain scaled by $s_\Tb$.
}
\label{fig:training_data-stress}
\end{figure}

\begin{figure}[h!]
\centering
\subfloat[][tension]
{\includegraphics[width=\figwidthtwo]{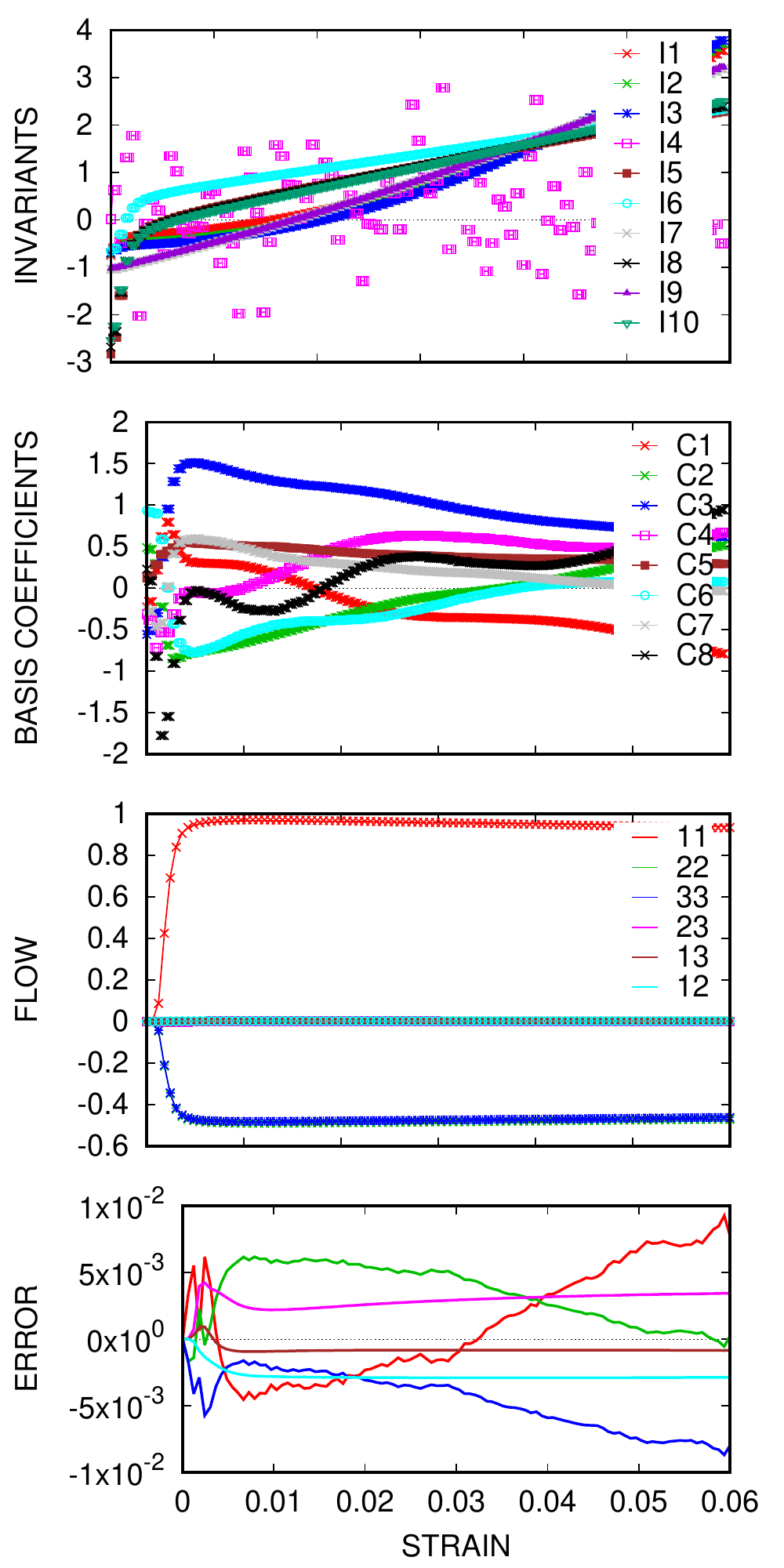}}
\subfloat[][shear]
{\includegraphics[width=\figwidthtwo]{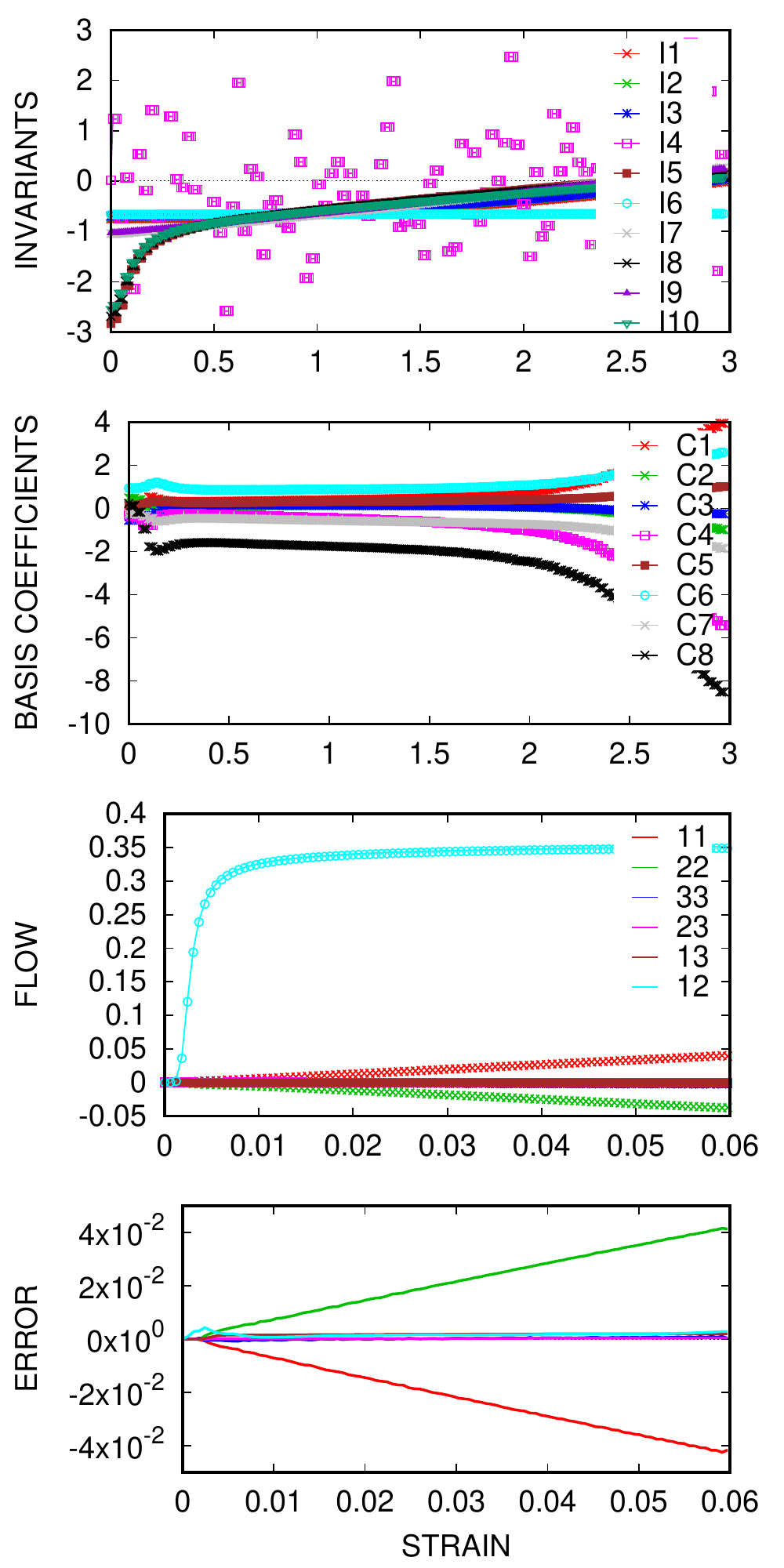}}
\caption{ 
Flow training data: (a) tension and (b) shear flow evolution with strain for CP model.
Top panels: scaled input invariants.
Second panels: scaled trained tensor basis coefficient functions.
Third panels: flow $\Db_p$ response (lines: model, points: data).
Bottom panel: error as function of strain scaled by $s_{\Db_p}$.
}
\label{fig:flow_training_data-flow}
\end{figure}

For this study we use data from the CP model to train the stress and flow TBNNs.
In particular, we collect data using 3 tension and 6 simple shear loading modes averaged over 570 random textures for each of the 10 polycrystalline realizations.
As mentioned in the Methods section, we give ourselves access to the (average) plastic state variables of the CP simulations and so we train the stress and flow TBNNs independently (and not simultaneously).
\fref{fig:training_data-stress} and \fref{fig:flow_training_data-flow} shows typical training data for the I3 stress and IF flow representations (refer to \tref{tab:stress_representations} and \tref{tab:flow_representations})  with $3\times 4$ and $5\times 8$ NNs, respectively.
The left column of \fref{fig:training_data-stress} and \fref{fig:flow_training_data-flow} show the tension response and the right columns show the shear response.
The upper panels show the (input) invariants and the (output) coefficient functions.
In general, the inputs and outputs are smooth and correlated, and all coefficient functions contribute.
The notable exception is the stress model in shear in which only the coefficient function of the linear basis element $\eb_e$ appears to contribute.
Note that all invariants are arguments to each coefficient function.
Note in \fref{fig:flow_training_data-flow} the zero invariant, $\tr \sigmab \equiv 0$, that becomes noise upon the input scaling described in \sref{sec:training}.
Apparently, the NN training learns to ignore this input since the outputs are smooth and regular.
The lower panels show: (a) the correspondence of the model (lines) and the data (points), and (b) the error as a function of strain.
The errors for each of the components are of comparable magnitude and tend to have an irregular pattern in the elastic region of the loading.
Note that with a C0 activation function (\eg, the Rectifying Unit $a(x) = \max(0,x)$) we observed distinct scallops and cusps in error curves (not shown for brevity).
Also it is remarkable that the errors of the flow model in shear are distinctly linear, which, perhaps, is related to the fact only the linear basis element is active.

\begin{figure}[h!]
\centering
\subfloat[][stress model]
{\includegraphics[width=\figwidthtwo]{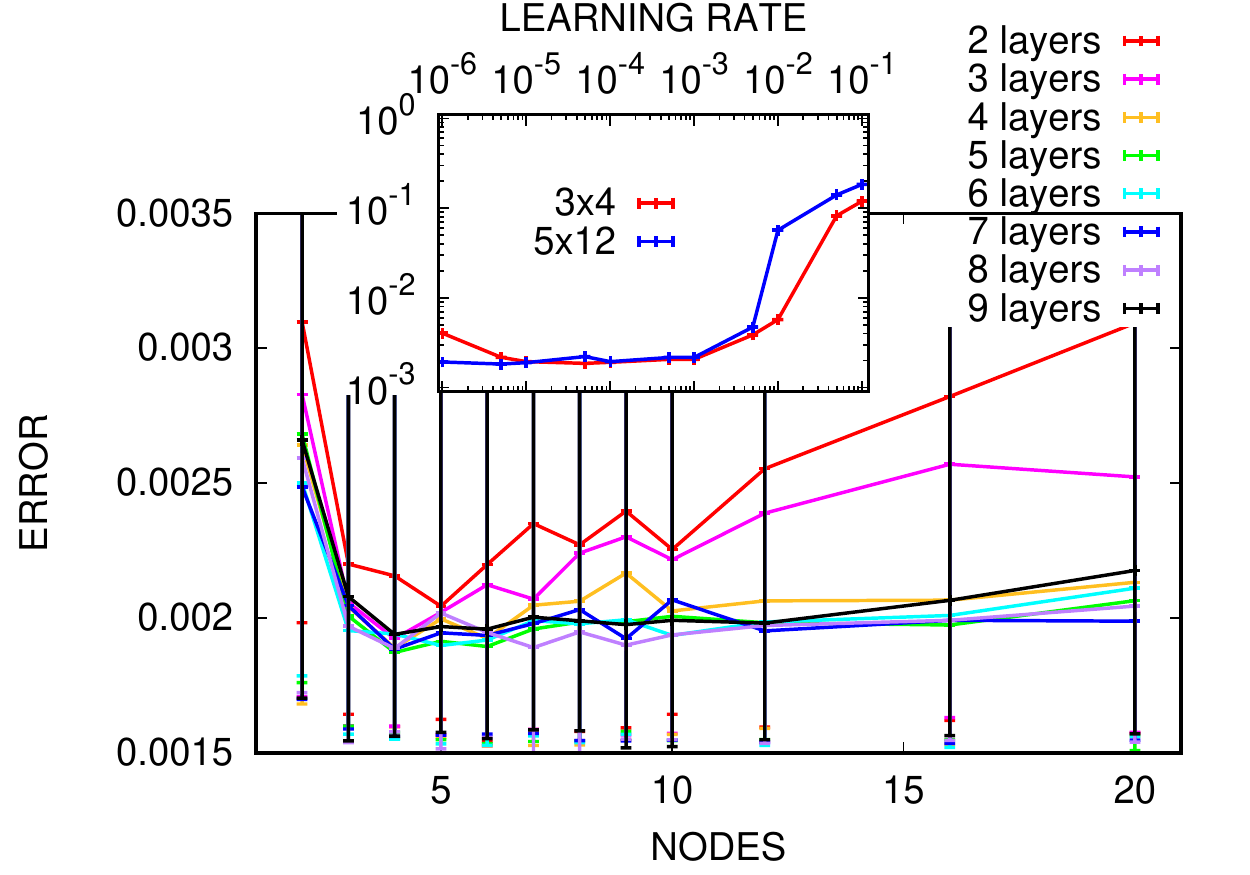}}
\subfloat[][flow model]
{\includegraphics[width=\figwidthtwo]{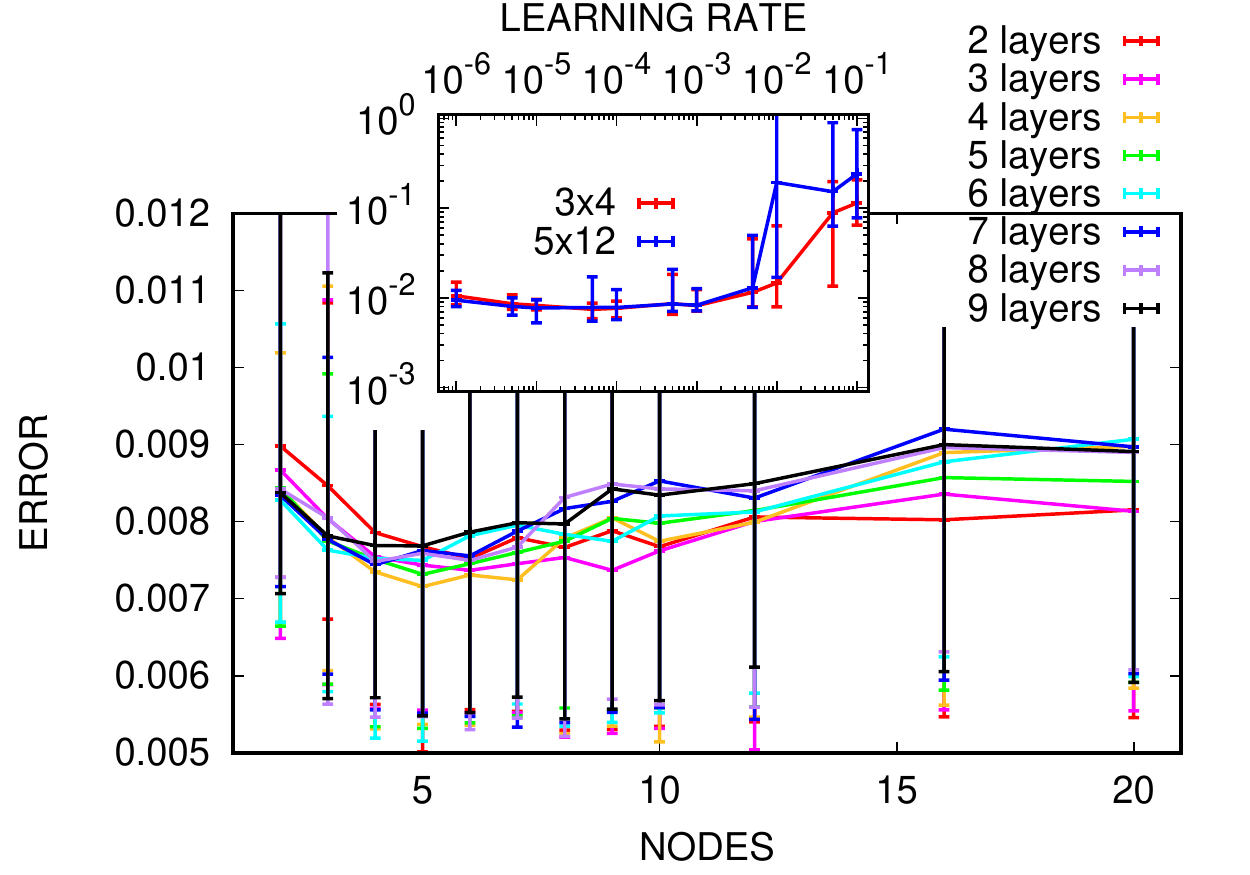}}
\caption{
Network optimization for 150 realizations of full representations for both stress and flow on CP data.
Training meta parameter for full representations for (I3) stress and (SF) flow on CP data.
Error bars denote min and max of error, and errors are scaled by $s_\Tb$ and $s_{\Db_p}$, respectively.
}
\label{fig:training_meta}
\end{figure}

These results are typical for a wide range of NN structures and (meta) training parameters.
\fref{fig:training_meta} shows the cross-validation errors of the stress and flow (scaled by $s_\Tb$ and $s_{\Db_p}$, respectively) using the full representations I3 and IF (refer to \tref{tab:stress_representations} and \tref{tab:flow_representations}, respectively).
As Schwartz-Ziv and Tishby \cite{shwartz2017opening} remark, trying to interpret the behavior of network from a single training tends to be meaningless; hence, we evaluate parametric and structural changes with an ensemble of at least 30 replicas models  $M_k \in \Mc$ in this and the following studies.
(The replicas are obtained by using different random seeds to produce the initial weights and thresholds.)
The insets show that the (initial) learning rate can have a strong effect on the errors, but once a small enough ($< 10^{-3}$) rate is selected the final errors are relatively insensitive to this parameter.
The main panels show the typical trends in errors ranging from under-representation (too small a network) to over-fitting (too large a network).%
\fnote{As mentioned, we require that in the training procedure that the error on the training $D$ and the testing $T$ data be comparable as failure to achieve parity in the errors is indicative of bad predictions and over-fitting.}
For the stress TBNN, $N=4$ nodes appears to be an optimum even for relatively shallow networks ($N < 4$) but the optimal number of nodes is relatively insensitive for $L>4$.
The flow TBNN shows analogous behavior but with a trade-off between nodes and layers, \eg for $L> 6$, $N=4$ appears to be best, while $N>4$ is better for shallower networks. 
These findings are somewhat obscured by the noise in the trend lines, which persists despite using the average of 150 replica networks.
Also, the convergence window (described in \sref{sec:training}) is an important meta parameter.
We obtained these results with a 4 iteration convergence window; a longer convergence window (\eg 10 iterations) shifts the best cross-validation to smaller networks (but also induces larger variance in error between replicas).
Since we want reliable error from each replica, we use a  4 iteration convergence window throughout the remainder of this work.
Lastly, we do not believe cross-validation is sufficient for determining completeness of network; however, these results indicate that optimal number of nodes is less than the number of input invariants for flow but greater than this matrix rank-based criterion for the stress representation.
Apparently, with respect to the training data, the NN is compressing the input for the flow network; and, hence, we conjecture that the NN is forming lower dimensional set of (alternate) invariants internally.

\begin{figure}[h!]
\centering
\subfloat[][stress]
{\includegraphics[width=\figwidthtwo,valign=c]{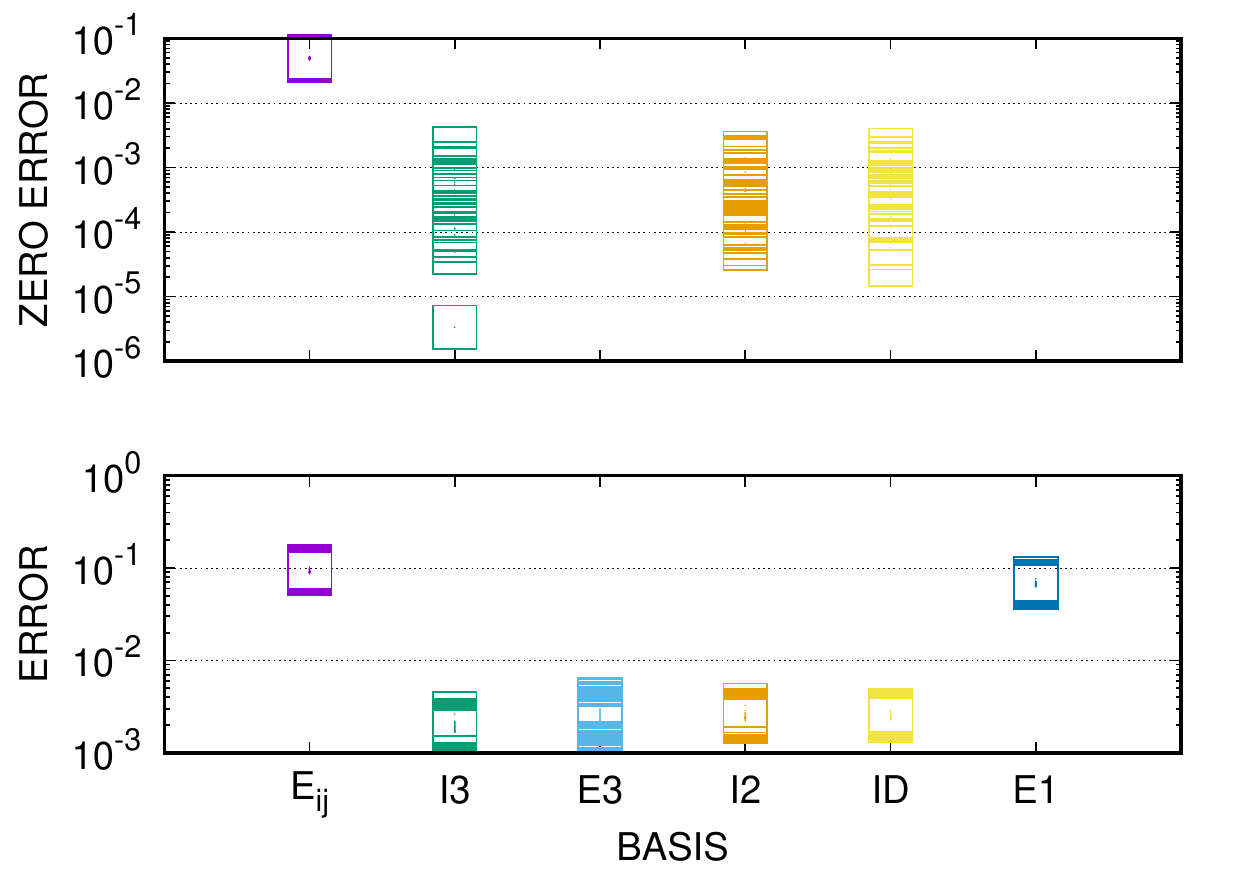}}
\subfloat[][flow]
{\includegraphics[width=\figwidthtwo,valign=c]{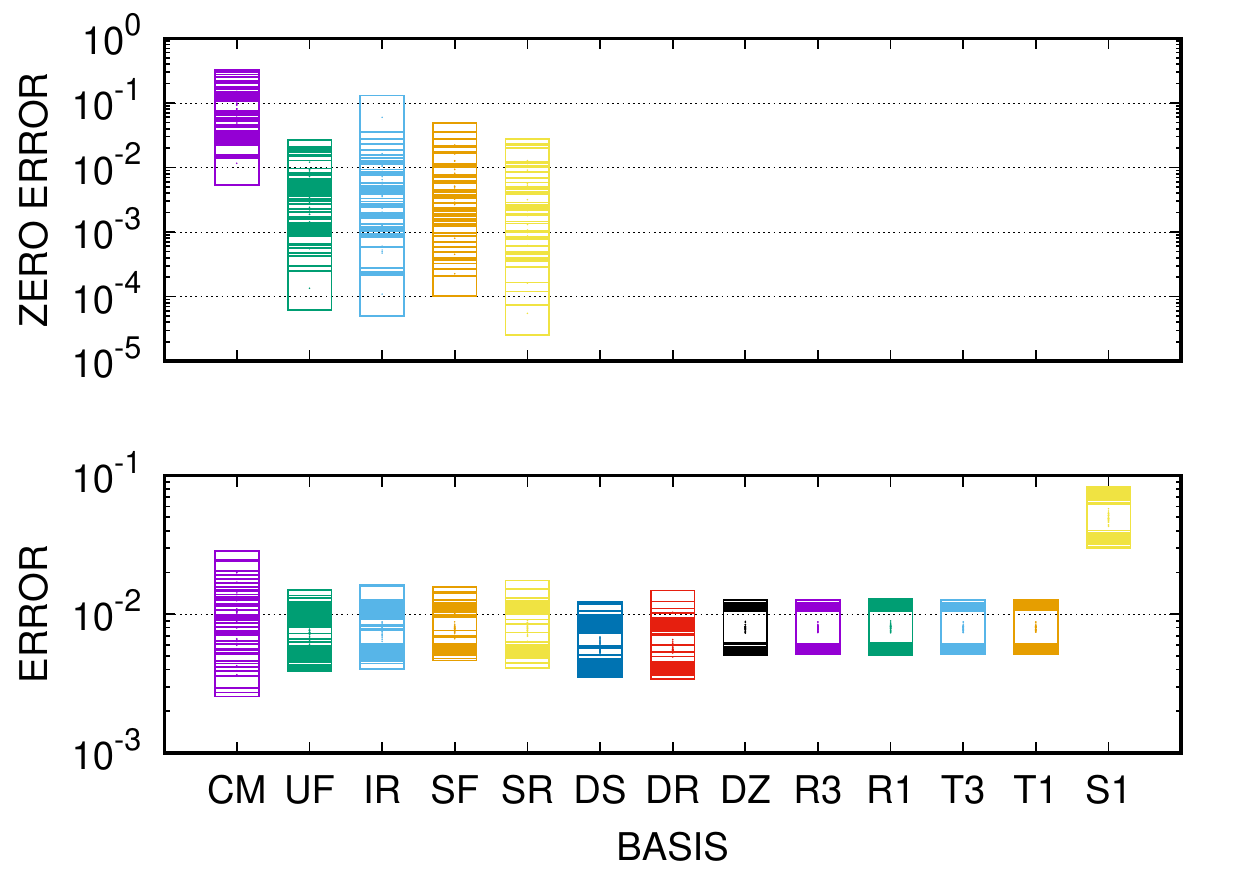}}
\caption{
Cross validation error as a function of representation for (a) stress,  (b) flow for CP data.
Errors are scaled by $s_\Tb$ and $s_{\Db_p}$, respectively.
Refer to \tref{tab:stress_representations} for stress representations and \tref{tab:flow_representations} for the flow representations.
}
\label{fig:representation}
\end{figure}

\fref{fig:representation} shows cross-validation error for the CP training data for various basis representation of the stress and flow functions (refer to \tref{tab:stress_representations} and \tref{tab:flow_representations} for the definition of the labels).
For the stress TBNNs, all (overall) errors are comparable with the exception of the component-based representation and the one term E1 representation (with tensor basis $\Bc = \{ \eb_e \}$).
Clearly, the E1 basis is not sufficient since it is akin to a one parameter Navier model of stress.
From the results of the two truncated bases, I2 and ID, it appears that correlated inputs, $\Bc = \{ \Ib, \eb_e\}$ (I2), train comparably to linearly independent inputs, $\{ \Ib, \dev \eb_e\}$ (ID, which uses a volumetric/deviatoric split). 
Also, the upper panel of \fref{fig:representation}a shows that the representations without embedded satisfaction of the zero-stress at zero-strain constraint, \eref{eq:zero_stress}, generally violate this constraint by about 1\% of the maximum stress.
For the flow TBNNs, all (overall) errors are comparable with the exception of the reduced scalar and tensor basis representation S1.
Also the other reduced representations (R3,R1,T3,T1) have slightly higher average errors than the full representations (UF,IF,IR,SF,DS,DS,DR,DZ) albeit with reduced variance.
The consistency of the representations with the zero-flow-at-zero-stress condition \eref{eq:zero_stress} generally follows whether powers of $\eb_e$ are included or not.
Clearly, cross-validation based on this limited dataset is not sufficient for decisive model selection but it does eliminate some representations.
By comparison, the component-based representations, $E_{ij}$ and CM, display higher errors, larger variance in the performance and poor zero-input-zero-output results.
Beyond the fundamentally different functional representation, these models are likely suffering from an insufficiency of data to learn the necessary properties  (the ones embedded in the generalized TBNN framework) accurately, as demonstrated in \cref{ling2016machine}.
Lastly, as discussed in the Theory section, we have embedded a number of properties in the representations, \eg symmetry, deviatoric flow, dissipation, and, generally, the violation of the learned properties is on par with what we illustrate with the zero-stress and zero-flow conditions.

In preliminary studies we also trained networks with different inputs and outputs. 
In general, the cross-validation errors were comparable over a variety of choices, for example using a symmetrized Mandel stress for the driving stress input to the flow rule.
We considered the rate $\dot{\overline{\Cb_p^{-1}}}$ of the inverse of the plastic right Cauchy-Green deformation tensor $\Cb_p = \Fb_p^T \Fb_p$ (as in Simo and Hughes \cite[Ch. 9]{Simo1998}) as the output of the flow rule and obtained similar cross-validation (and prediction) performance.
Also noteworthy, we employed both the elastic and the full left Cauchy-Green stretch tensors as history inputs and the full Cauchy stress as a driving stress input.
These inputs resulted in similar cross-validation except when we paired the elastic Cauchy-Green stretch with the highly correlated Cauchy stress we observed slightly higher errors (and less variance among the errors).

\begin{figure}[h!]
\centering
\subfloat[][tension]
{\includegraphics[width=\figwidthtwo]{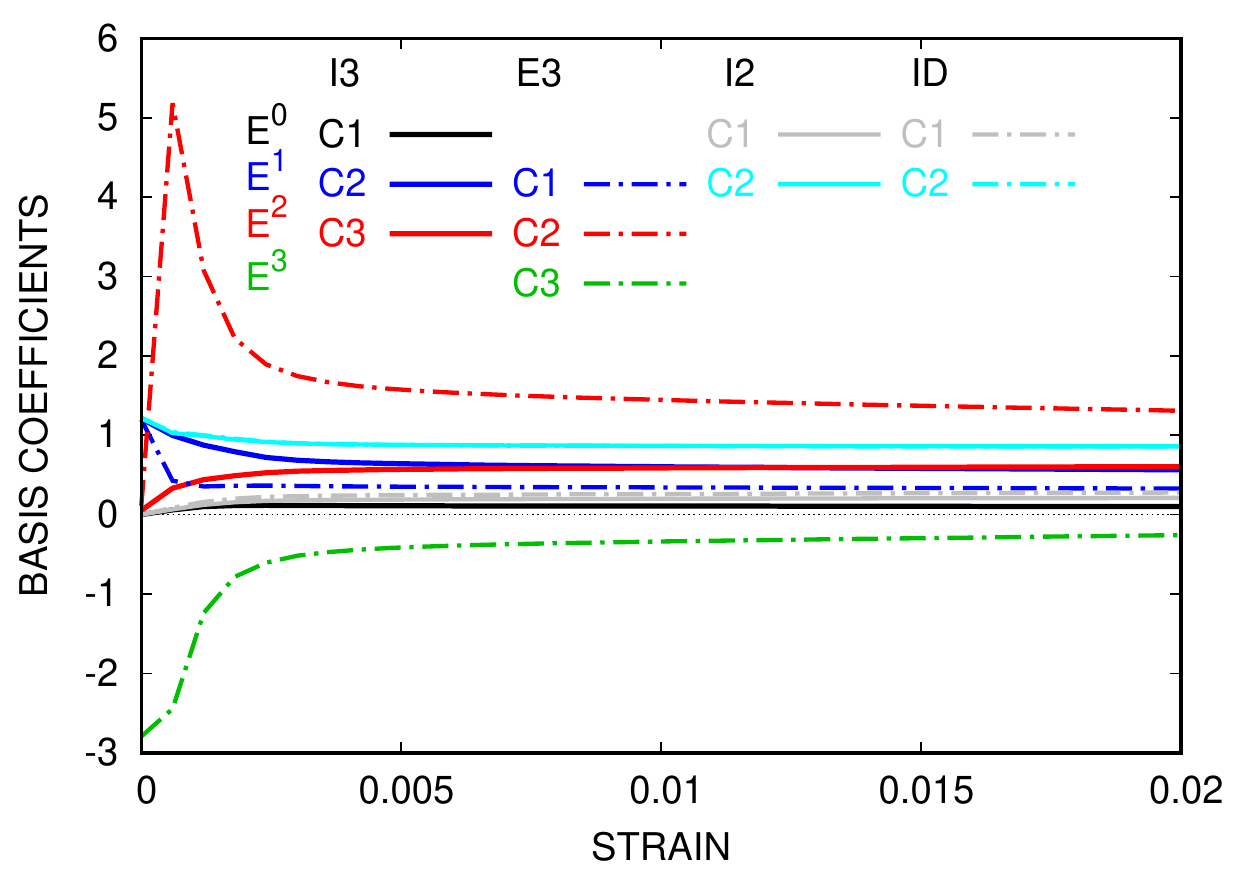}}
\subfloat[][shear]
{\includegraphics[width=\figwidthtwo]{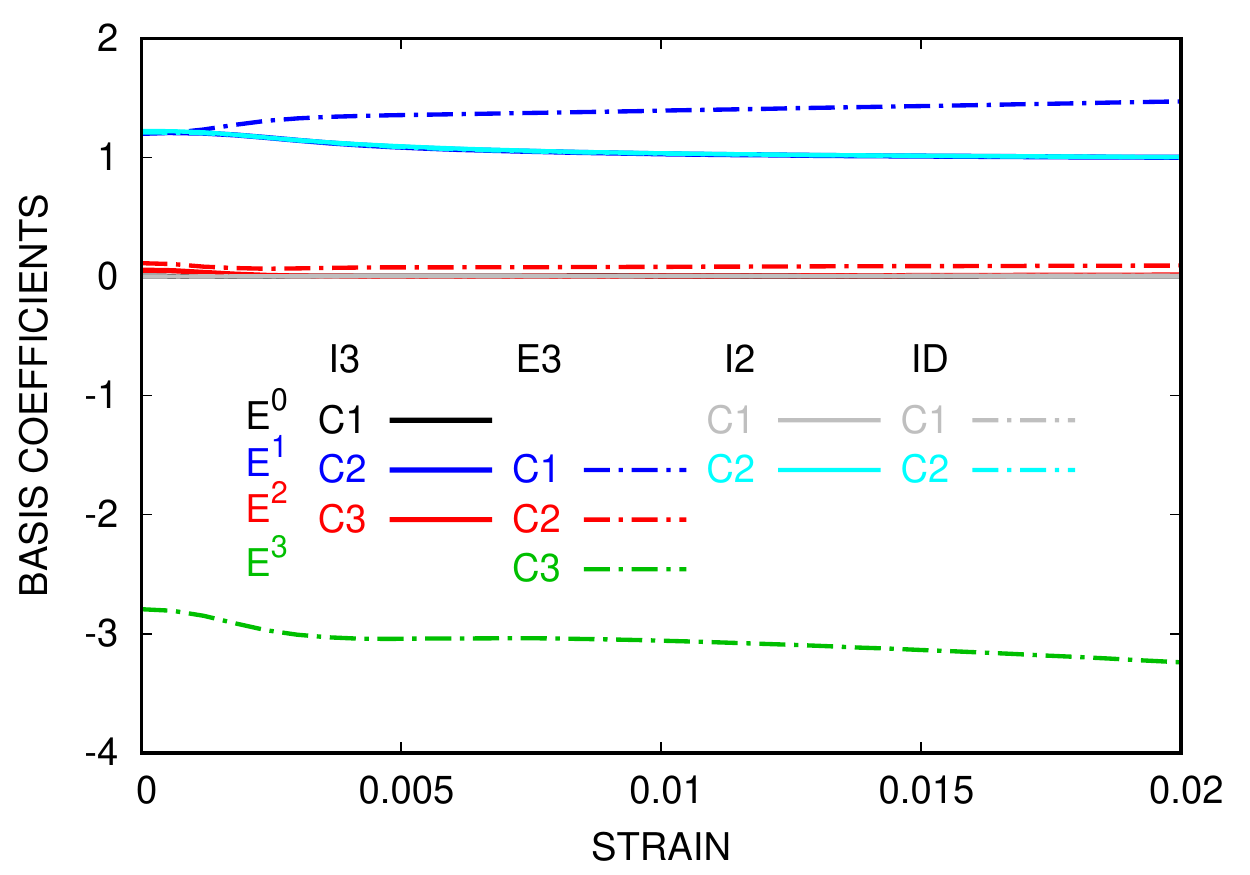}}

\caption{
Stress tensor basis coefficients in shear and tension using different bases.
}
\label{fig:stress_coefs}
\end{figure}

\begin{figure}[h!]
\centering

\subfloat[][tension]
{\includegraphics[width=\figwidthtwo]{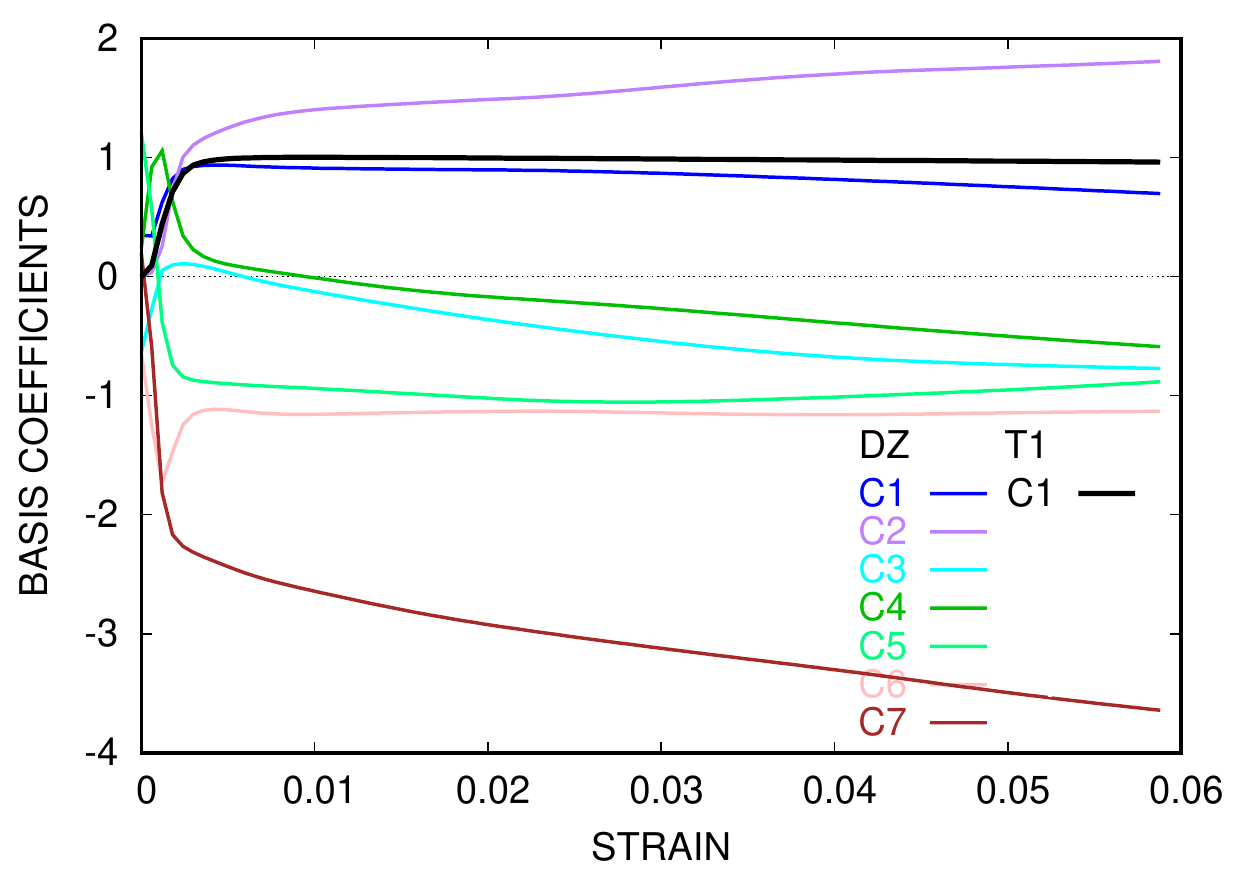}}
\subfloat[][shear]
{\includegraphics[width=\figwidthtwo]{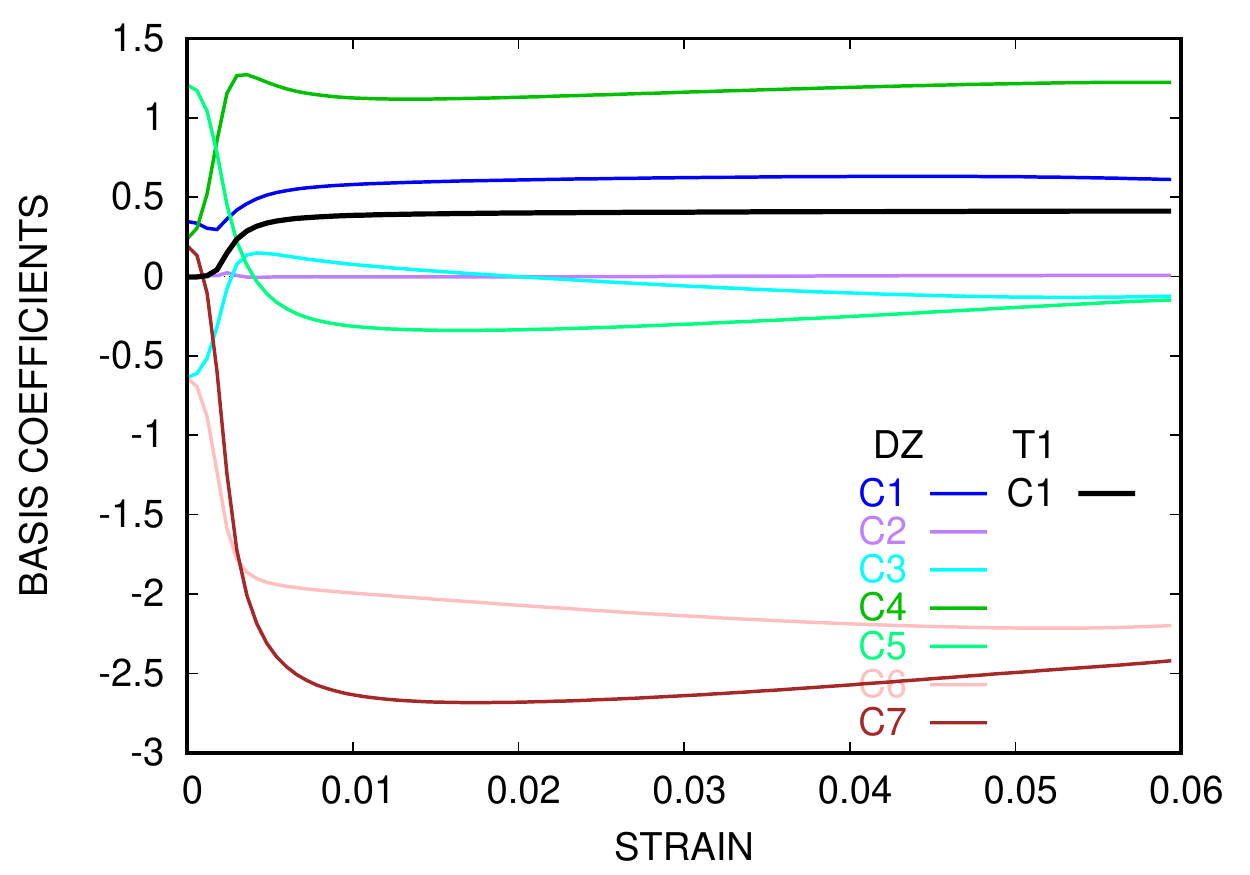}}
\caption{
Flow Tensor basis coefficients in shear and tension using DZ and T1 bases.
}
\label{fig:flow_coefs}
\end{figure}

\fref{fig:stress_coefs} and \fref{fig:flow_coefs} show the response of the NN coefficient functions to tension and shear, for stress and flow, respectively.
In these plots, each coefficient is scaled according to \eref{eq:scaling} so that coefficient functions of higher order terms can be plotted on par with those of lower order terms.
First, we notice that the E3 basis achieves zero-stress at zero-strain satisfaction exactly at the expense of a more complex, larger magnitude per component response than I3, as the higher order term $\eb_e^3$ apparently needs compensation by the component functions, refer to \eref{eq:spectral_components}.
Also, we see more evidence that the truncated representations, I2 and ID, have almost indistinguishable response despite ID having a linearly independent tensor basis.
For the flow representation we only compare the DZ and T1 representations for clarity.
Note that T1 is much simpler in form (one tensor basis element versus ten) than DR, which has a complete basis, and its response is simpler while achieving comparable cross-validation error to DZ.
Also evident from both tension and shear response, DR builds a similar response to T1 by letting all/most components contribute.
Also significant, the coefficient of $\sigmab^2$, C2, is essentially zero throughout the shear trajectory but not the tension, which implies that the NN may not be learning dissipation is an important property.
This is in contrast with the DR representation (not shown) where the corresponding coefficient is essentially zero for both tension and shear.
For both the stress and flow models, the coefficient responses generally resemble the trends in the stress and flow data, with large changes up to the elastic-plastic transition at strain $> 0.002$ followed by relatively constant values in the plastic regime.
This is consistent with the expectation that in fully developed plastic flow (in a constant direction with negligible hardening) the elastic state and the plastic flow are constant.

\begin{table}
\centering
\footnotesize
\begin{tabular}{|c|c|c|}
\hline
         & scalar   & tensor \\
\hline
\hline
E$_{ij}$ & component $\left[\eb_e\right]_{ij}$ & component $\eb_i \otimes \eb_j$ \\
\hline
I3       & full $\{ \tr \eb_e, \tr \eb_e^2, \tr \eb_e^3 \}$ & full $\{ \Ib,\eb_e,\eb_e^2 \}$  \\
\hline
E3       & full $\{ \tr \eb_e, \tr \eb_e^2, \tr \eb_e^3 \}$ & full, shifted $\{ \eb_e,\eb_e^2,\eb_e^3 \}$  \\
\hline
I2       & full $\{ \tr \eb_e, \tr \eb_e^2, \tr \eb_e^3 \}$ & reduced $\{ \Ib,\eb_e \}$  \\
\hline
ID       & full $\{ \tr \eb_e, \tr \eb_e^2, \tr \eb_e^3 \}$ & reduced, independent $\{ \Ib,\dev\eb_e \}$  \\
\hline
E1       & full $\{ \tr \eb_e, \tr \eb_e^2, \tr \eb_e^3 \}$ & reduced $\{ \eb_e \}$  \\
\hline
\end{tabular}
\caption{Stress representations
}
\label{tab:stress_representations}
\end{table}

\begin{table}
\centering
\footnotesize
\begin{tabular}{|c|c|}
\hline
CM & component basis \\
\hline
UF & full: $\{ \tr \bb, \tr \bb^2, \tr \bb^3, \tr \sb, \tr \sb^2, \tr \sb^3, \tr \sb \bb, \tr \sb \bb^2, \tr \sb^2 \bb, \tr \sb^2 \bb^2 \}$  \\
   & unsymmetric,full: $\{ \Ib, \bb, \sb, \bb^2, \sb^2, \bb \sb,   \sb^2 \bb, \bb \sb^2 \}$ \\
\hline
IF & full: $\{ \tr \bb, \tr \bb^2, \tr \bb^3, \tr \sb, \tr \sb^2, \tr \sb^3, \tr \sb \bb, \tr \sb \bb^2, \tr \sb^2 \bb, \tr \sb^2 \bb^2 \}$  \\
   & full: $\{ \Ib, \bb, \sb, \bb^2, \sb^2, \sym \bb \sb,   \sym \sb^2 \bb, \sym \bb \sb^2 \}$ \\
\hline
IR & no $\tr \sb$: $\{ \tr \bb, \tr \bb^2, \tr \bb^3, \tr \sb, \tr \sb^2, \tr \sb^3, \tr \sb \bb, \tr \sb \bb^2, \tr \sb^2 \bb, \tr \sb^2 \bb^2 \}$  \\
   & full: $\{ \Ib, \bb, \sb, \bb^2, \sb^2, \sym \bb \sb,   \sym \sb^2 \bb, \sym \bb \sb^2 \}$ \\
\hline
SF & full $\{ \tr \bb, \tr \bb^2, \tr \bb^3, \tr \sb, \tr \sb^2, \tr \sb^3, \tr \sb \bb, \tr \sb \bb^2, \tr \sb^2 \bb, \tr \sb^2 \bb2 \}$  \\
   & shifted, full $\{ \bb, \sb, \bb^2, \sb^2, \sym \bb \sb, \bb^3,   \sym \sb^2 \bb, \sym \bb \sb^2 \}$ \\
\hline
SR & no $\tr \sb$: $\{ \tr \bb, \tr \bb^2, \tr \bb^3, \tr \sb, \tr \sb^2, \tr \sb^3, \tr \sb \bb, \tr \sb \bb^2, \tr \sb^2 \bb, \tr \sb^2 \bb2 \}$  \\
   & shifted,full: $\{ \bb, \sb, \bb^2, \sb^2, \sym \bb \sb, \bb^3,   \sym \sb^2 \bb, \sym \bb \sb^2 \}$ \\
\hline
DS & full: $\{ \tr \bb, \tr \bb^2, \tr \bb^3, \tr \sb, \tr \sb^2, \tr \sb^3, \tr \sb \bb, \tr \sb \bb^2, \tr \sb^2 \bb, \tr \sb^2 \bb2 \}$  \\
   & deviatoric,shifted,full: $\{ \dev \bb, \dev \sb, \dev \bb^2, \dev \sb^2, \dev \sym \bb \sb, \dev \sb^3,  \dev \sym \sb^2 \bb, \dev \sym \bb \sb^2 \}$ \\
\hline
DR & no $\tr \sb$: $\{ \tr \bb, \tr \bb^2, \tr \bb^3, \tr \sb^2, \tr \sb^3, \tr \sb \bb, \tr \sb \bb^2, \tr \sb^2 \bb, \tr \sb^2 \bb2 \}$  \\
   & deviatoric,shifted,full: $\{ \dev \bb, \dev \bb^2, \dev \sb, \dev \sb^2, \dev \sb^3, \dev \sym \bb \sb,  \dev  \sym \sb^2 \bb, \dev \sym \bb \sb^2 \}$ \\
\hline
DZ & no $\tr \sb$: $\{ \tr \bb, \tr \bb^2, \tr \bb^3, \tr \sb^2, \tr \sb^3, \tr \sb \bb, \tr \sb \bb^2, \tr \sb^2 \bb, \tr \sb^2 \bb2 \}$  \\
   & deviatoric,dissipative: $\{ \dev \sb, \dev \sb^2, \dev \sb^3, \dev \sym \bb \sb, \dev \sym \sb^2 \bb, \dev \sym \bb \sb^2 \}$ \\
\hline
R3 & full: $\{ \tr \bb, \tr \bb^2, \tr \bb^3, \tr \sb, \tr \sb^2, \tr \sb^3, \tr \sb \bb, \tr \sb \bb^2, \tr \sb^2 \bb, \tr \sb^2 \bb2 \}$  \\
   & reduced, dissipative: $\{ \sb, \dev \sb^3 \}$ \\
\hline
R1 & full: $\{ \tr \bb, \tr \bb^2, \tr \bb^3, \tr \sb, \tr \sb^2, \tr \sb^3, \tr \sb \bb, \tr \sb \bb^2, \tr \sb^2 \bb, \tr \sb^2 \bb2 \}$  \\
   & reduced, dissipative: $\{ \sb \}$ \\
\hline
T3 & no $\tr \sb$: $\{ \tr \bb, \tr \bb^2, \tr \bb^3, \tr \sb^2, \tr \sb^3, \tr \sb \bb, \tr \sb^2 \bb, \tr \sb \bb^2, \tr \sb^2 \bb2 \}$  \\
   & reduced, dissipative: $\{ \sb, \sb^3 \}$ \\
\hline
T1 & no $\tr \sb$: $\{ \tr \bb, \tr \bb^2, \tr \bb^3, \tr \sb^2, \tr \sb^3, \tr \sb \bb, \tr \sb^2 \bb, \tr \sb \bb^2, \tr \sb^2 \bb2 \}$  \\
   & reduced, dissipative: $\{ \sb \}$ \\
\hline
S1 & reduced: $\{ \tr \bb \}$  \\
   & reduced, dissipative: $\{ \sb \}$ \\
\hline
\end{tabular}
\caption{Flow representations
sym
has $\Ib$ (zero error)
is $\dev$
has $\tr \sigmab$ (noise invariant)
has $\bb_p$ in tensor basis
full scalar basis
}
\label{tab:flow_representations}
\end{table}

\subsection{Validation}

Our validations studies include tests of: (a) completeness of representation and training data, and (b) robustness to perturbation/continuity.

\begin{figure}[h!]
\centering

\subfloat[][11-stress, component]
{\includegraphics[width=1.2\figwidthtwo]{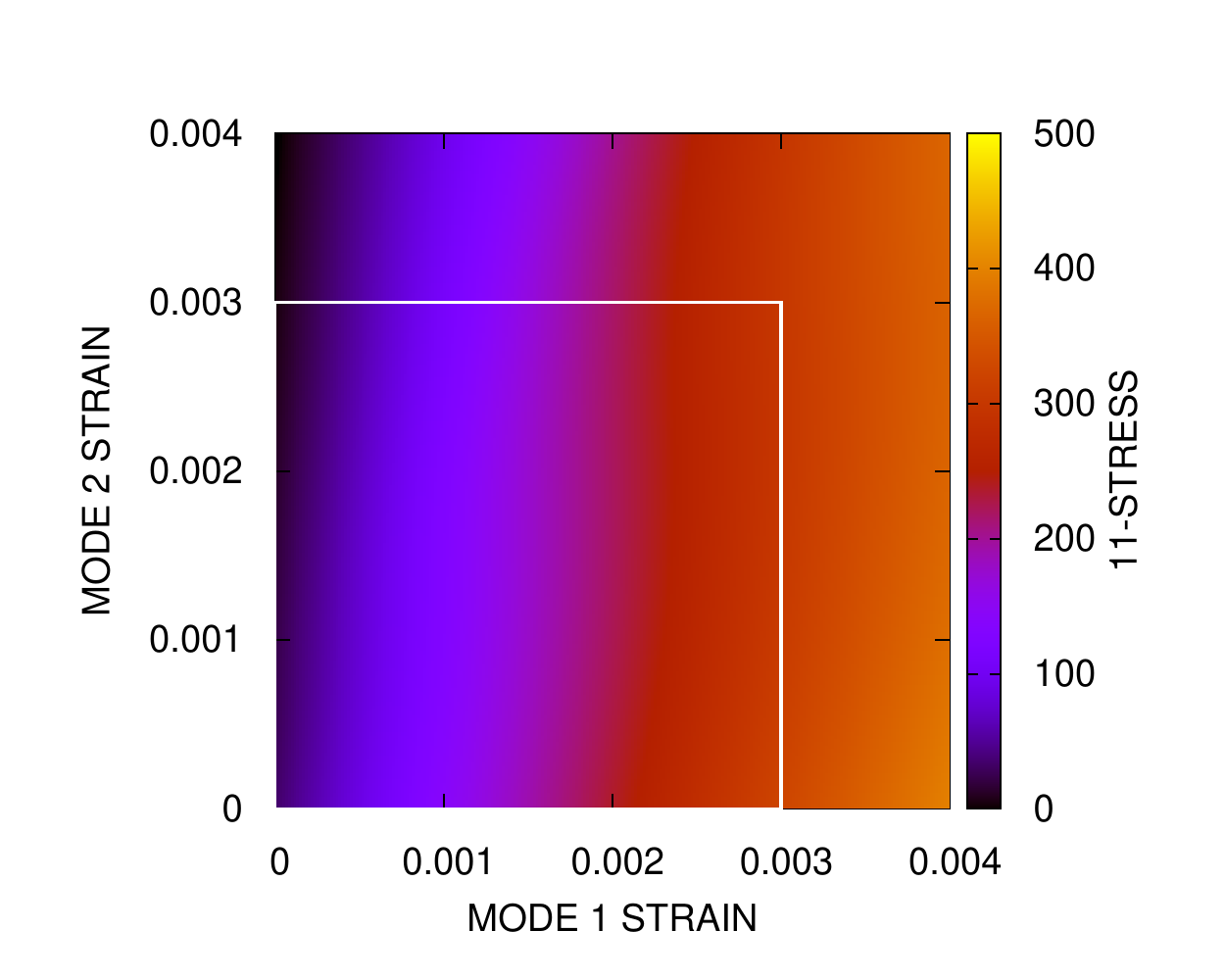}}
\subfloat[][22-stress, component]
{\includegraphics[width=1.2\figwidthtwo]{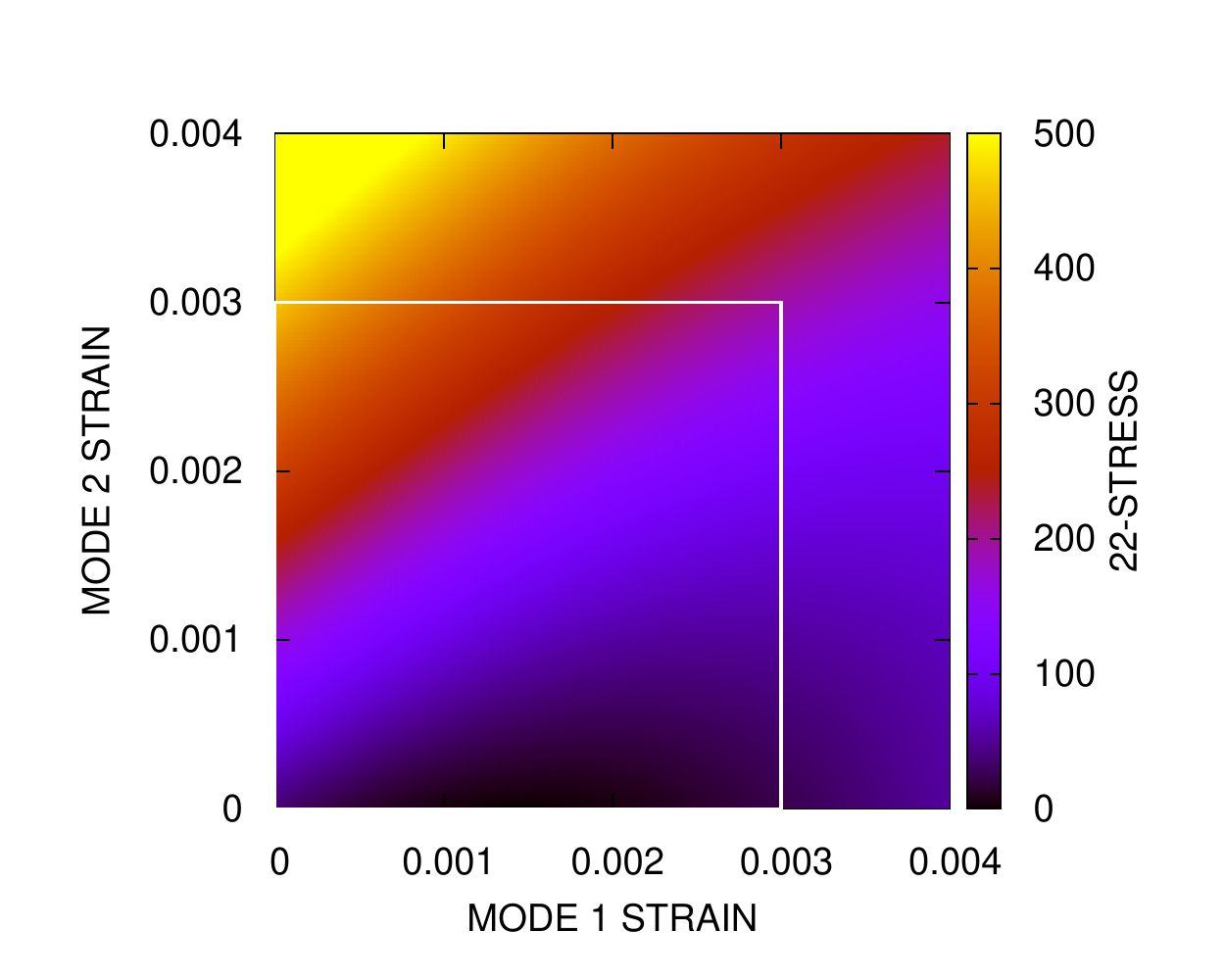}}

\subfloat[][11-stress, crystal]
{\includegraphics[width=1.2\figwidthtwo]{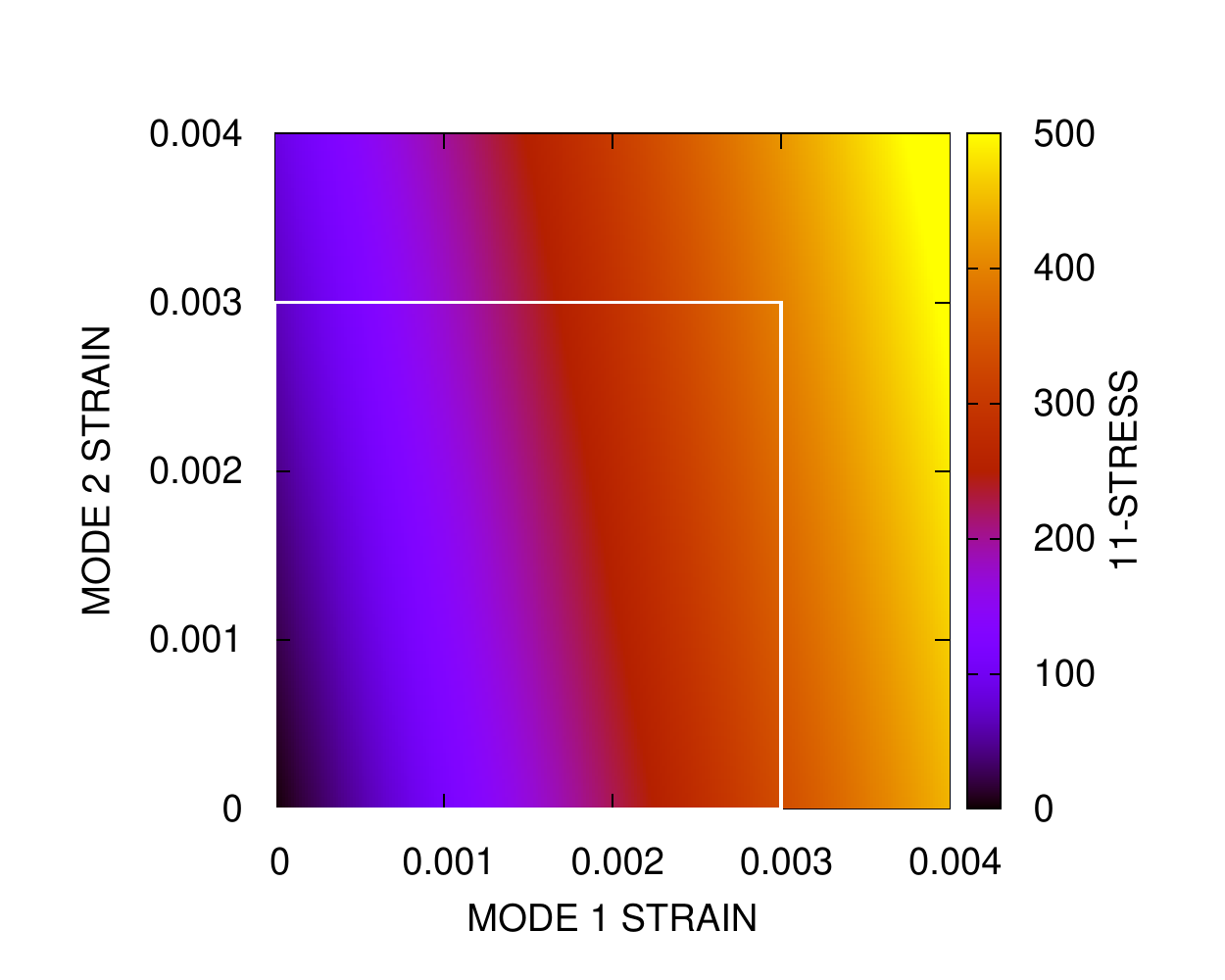}}
\subfloat[][11-stress, TBNN]
{\includegraphics[width=1.2\figwidthtwo]{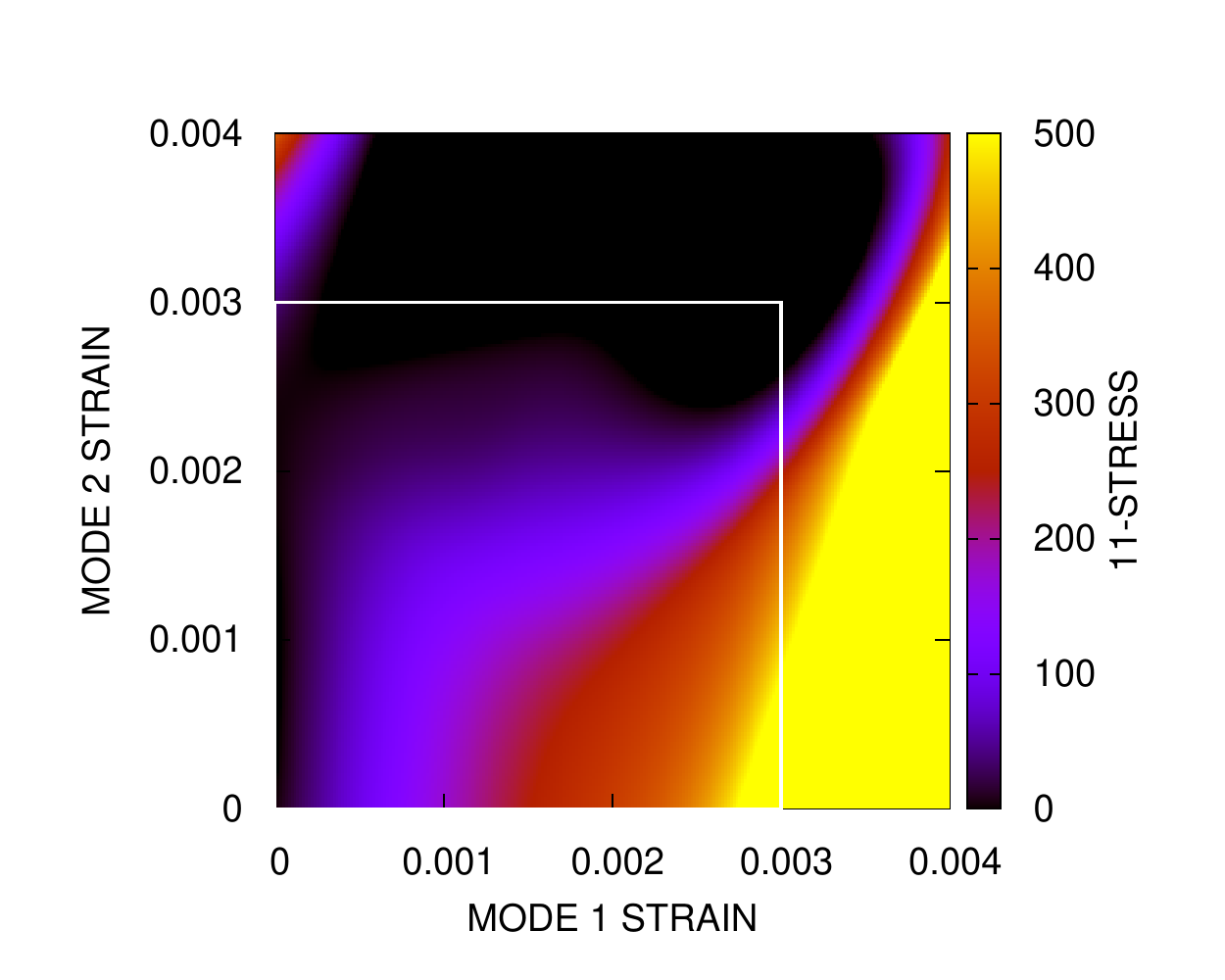}}

\caption{ 
Stress response of (a),(b) component $E_{ij}$, (c) underlying crystal, and (d) TBNN E3 models to bimodal stretch.
The TBNN 11-stress (c) and 22-stress (not shown) responses are symmetric across the diagonal.
Note white box outlines limit of training data which lies along the axes.
}
\label{fig:biaxial_stress}
\end{figure}

As we already have indications that a training set composed of only tension and shear may be insufficient, we computed the (E3) TBNN and ($E_{ij}$) component-based NN stress models' response to bimodal stretch  $\eb_e(\epsilon_1,\epsilon_2) = \epsilon_{1} \Mb_1 + \epsilon_{2} \Mb_2$, where the modes are the (tension) training modes $\Mb_i(\epsilon) = \epsilon \left( \eb_i \otimes \eb_i - \nu \left(\Ib - \eb_i \otimes \eb_i \right) \right)$.
Hence, the (tension) training data aligns with axes and here the models represent the data well.
\fref{fig:biaxial_stress} shows that the model responses are significantly different away from the training data, the limits of which are denoted by the white box outline on the contour plots.
In particular, the component model does not give a symmetric response, compare \fref{fig:biaxial_stress}a and \fref{fig:biaxial_stress}b, which is to be expected since each component is independently (and imperfectly) trained.
Both models have regions of negative stress and large stresses outside the training limits. 
\fref{fig:biaxial_stress}c shows expectation for the stress from the crystal elasticity underlying the response of the polycrystalline aggregate.
The TBNN response has the most complex trends likely due to its formulation on invariants and the 11-stress response of the component model arguably represents the expected stress best, albeit with a distinct error in the gradient.
This result illustrates that acceptable cross-validation errors along limited training data does not necessarily lead to comparably acceptable interpolation, nor extrapolation, with NNs.

\begin{figure}[h!]
\centering
\subfloat[][ stress: training error]
{\includegraphics[width=\figwidthtwo]{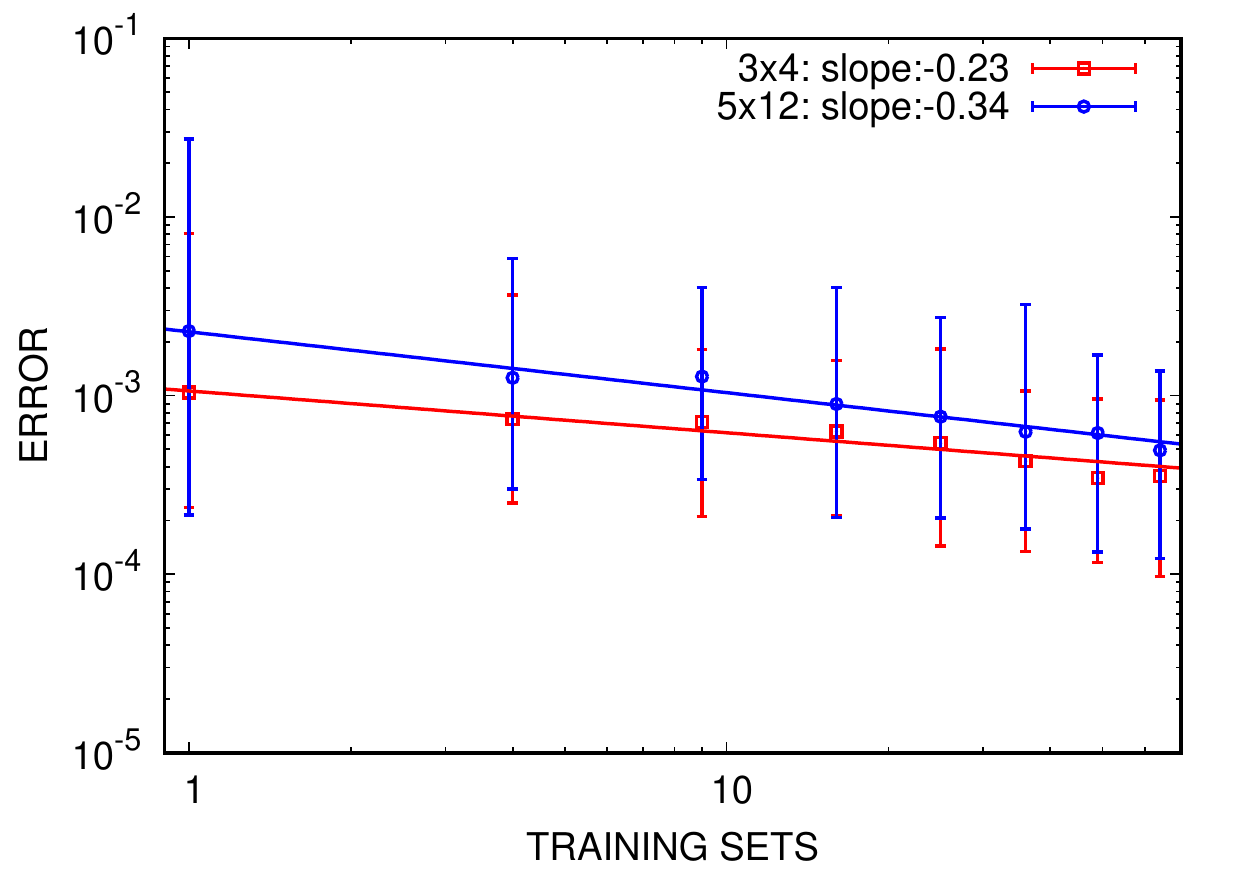}}
\subfloat[][ flow: training error]
{\includegraphics[width=\figwidthtwo]{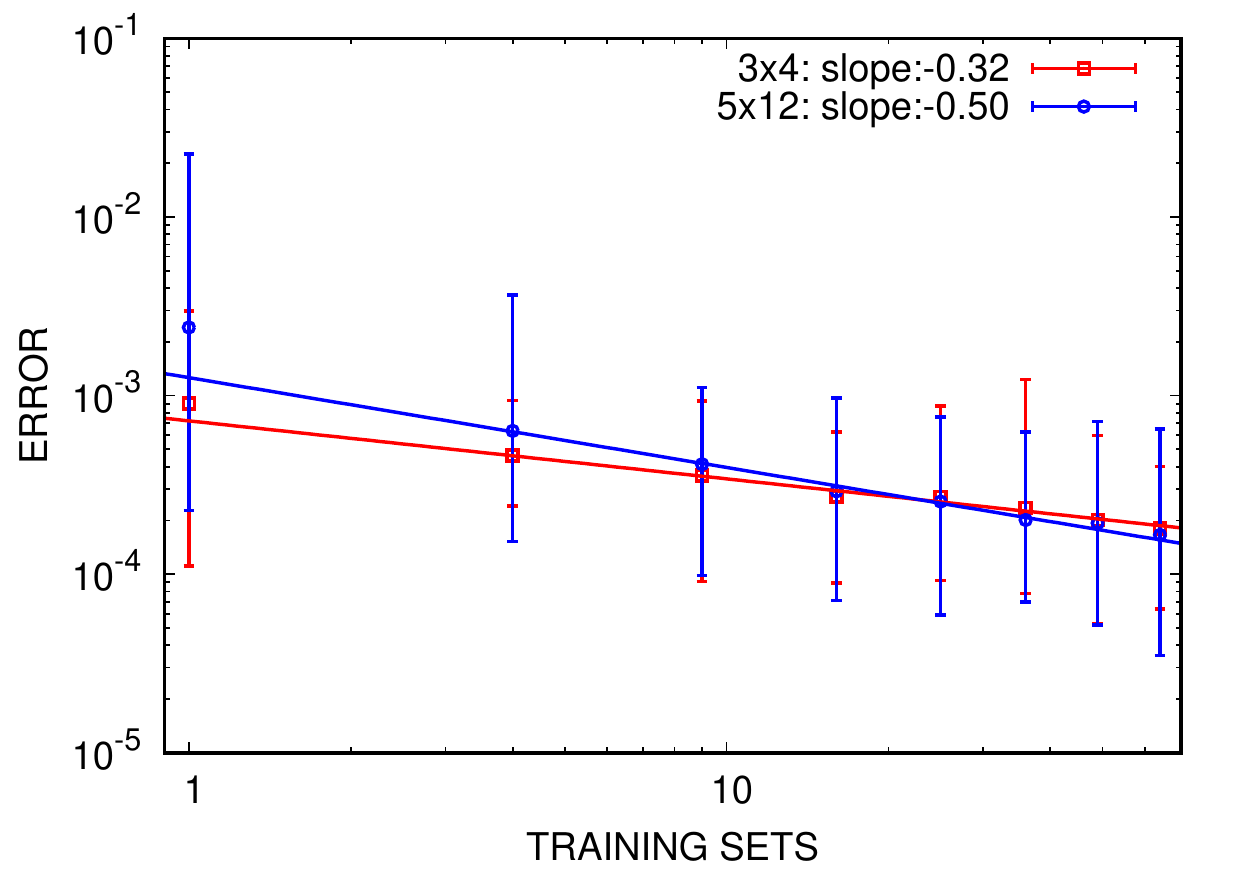}}

\subfloat[][ stress: K-L divergence]
{\includegraphics[width=\figwidthtwo]{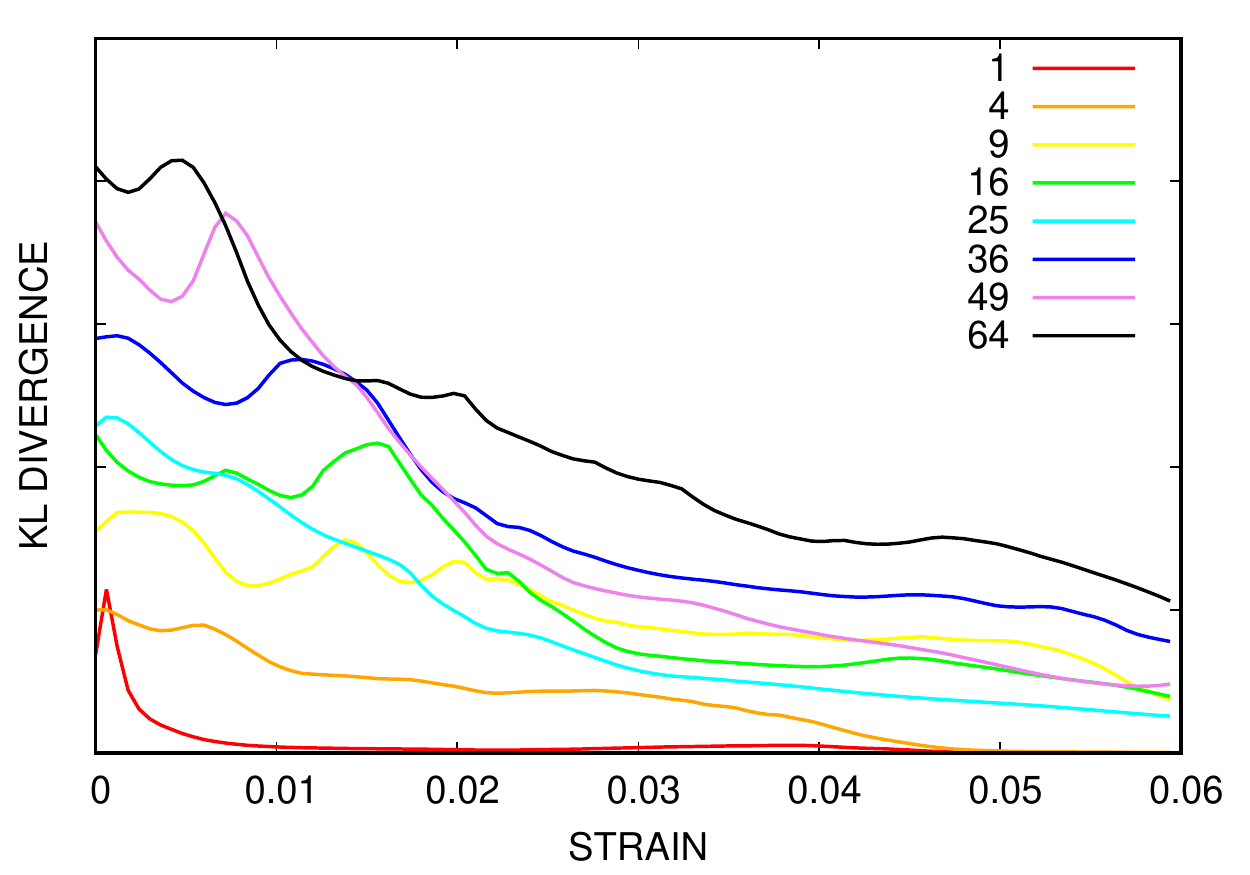}}
\subfloat[][ flow: K-L divergence]
{\includegraphics[width=\figwidthtwo]{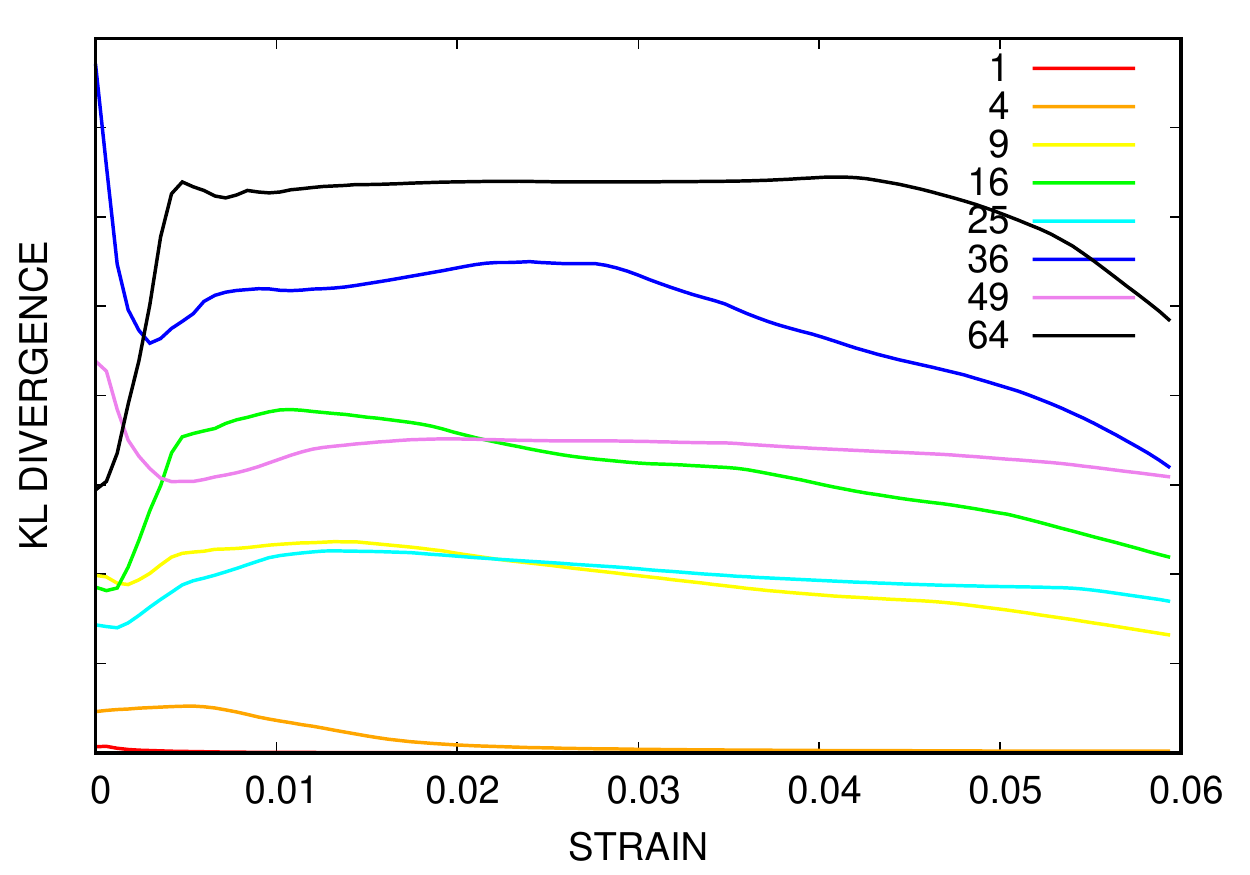}}
\caption{
Error as a function of training data span and network size (a,b), and 
model information content relative to an uninformed/untrained model (c,d). 
Note each training set/trajectory has 100 state samples from the VP (known underlying) model and the convergence rates are reported in terms of number training sets (not number of state samples).
Also, the error bars reflect the variance in the training errors across the ensemble of NN.
}
\label{fig:data_sufficiency}
\end{figure}

To further investigate how much data and what variety of data is needed sufficiently train the constitutive models we employed the simple VP underlying model to generate data for loading modes that are symmetric and monotonic: $\Fb(t) = t \sum_{i=1}^3 \lambda_i \eb_i \otimes \eb_i$.
In particular, the training datasets $D_n$ are comprised of $n$ trajectories with 100 state points each.
The ${\lambda_i}$ for each trajectory were uniformly sampled on the 2-sphere (using minimum energy points \cite{sloan2004extremal}).%
\fnote{Note the uniform sampling points nest, in the sense that a larger set $D_m$ contains all the points of a smaller set $D_n$, $m>n$.}
The testing dataset $T$ consisted of 10 trajectories given by random samples on the 2-sphere.
\fref{fig:data_sufficiency} shows the change in accuracy of the model relative to the random sample test $T$ set and the information gain with increasing the size of the training $D_n$ dataset.
The decreasing errors in \fref{fig:data_sufficiency}a,b  with more training suggests completeness of the representation, and the slightly higher rate of convergence for the larger network indicates the complexity of the underlying function.
Also the variability of the models is decreasing with more data, which gives context for the variability of the models trained only with the CP data.
The decrease rate is modest ($n^{-a}$, where $a\in [0.2,0.5]$ and $n$ is the number of trajectories that each contain 100 points) but the variability in response is also decreasing with more data.

To measure of how much information has been gained by training the NN (relative to its untrained state), we use the Kullback-Leibler (KL) divergence:
\begin{equation}\label{eq:kldiv}
g_j(D_n) = \int p (T | D_n) \log \frac{ p (T | D_n)} {p (T)} \mathrm{d}\ys_j
\end{equation}
evaluated with the assistance of standard kernel density estimators.
Here $D_n$ is a training set, $T$ is the independent test data,
and $p (T | D_n)$ is the probability density function (PDF) of the predictions $\ys_j$ using an ensemble of models $M_k \in \Mc$ and the (fixed) data inputs $\xs_j$, $j$ indexes the state and prescribed strain, and $p (T) \equiv p(T | D_0 )$.
In \fref{fig:data_sufficiency}c,d  we see that: (a) both the stress and flow models are steadily differentiating themselves from their untrained state with increased training data, (b) the largest changes appear to occur in the initial increases in training data and yet convergence of the KL divergence is not reached even with $D_{64}$, and (c) the stress is gaining more information from the low strain data and the flow model is gaining the most information from the post-yield data, which is physically intuitive.

\begin{figure}[h!]
\centering

\subfloat[][variance:stress]
{\includegraphics[width=\figwidthtwo]{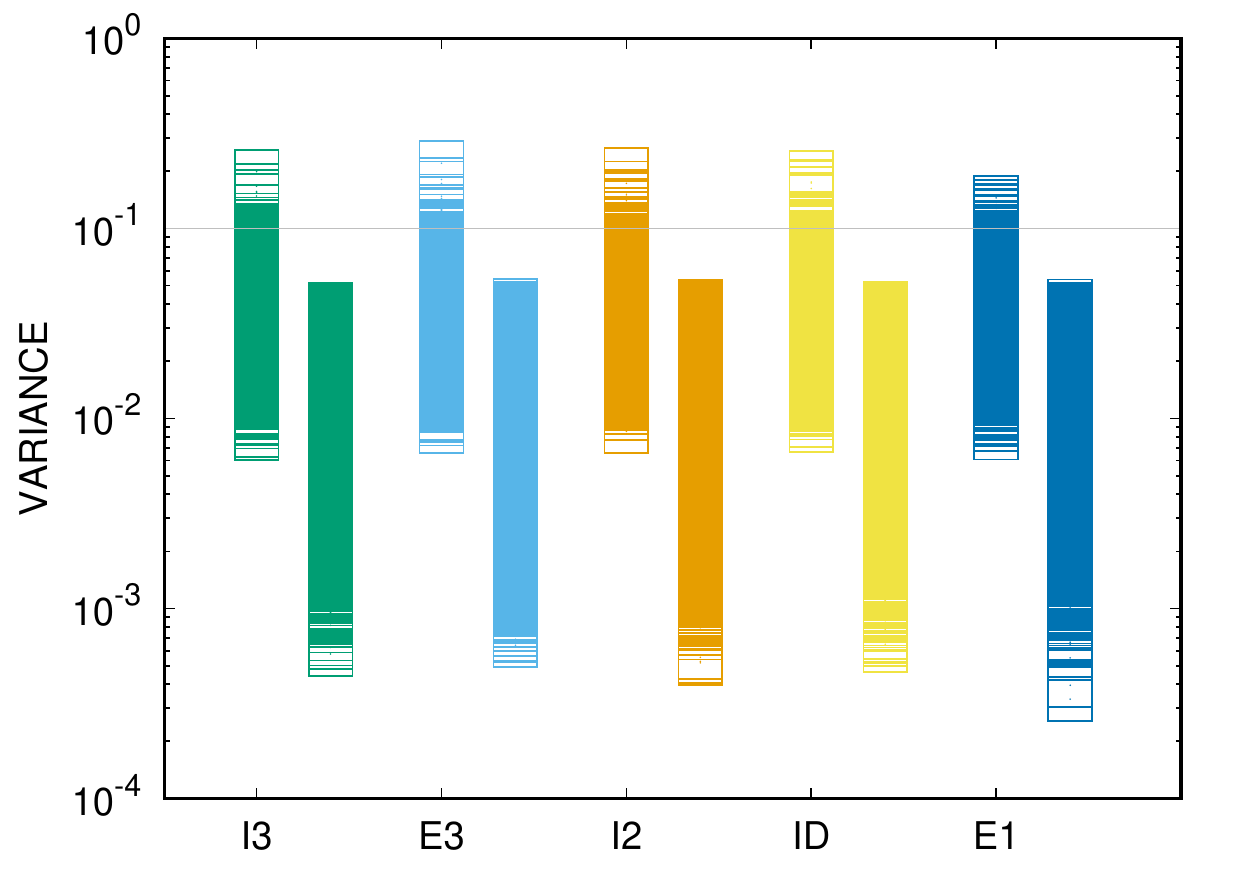}}
\subfloat[][variance:flow]
{\includegraphics[width=\figwidthtwo]{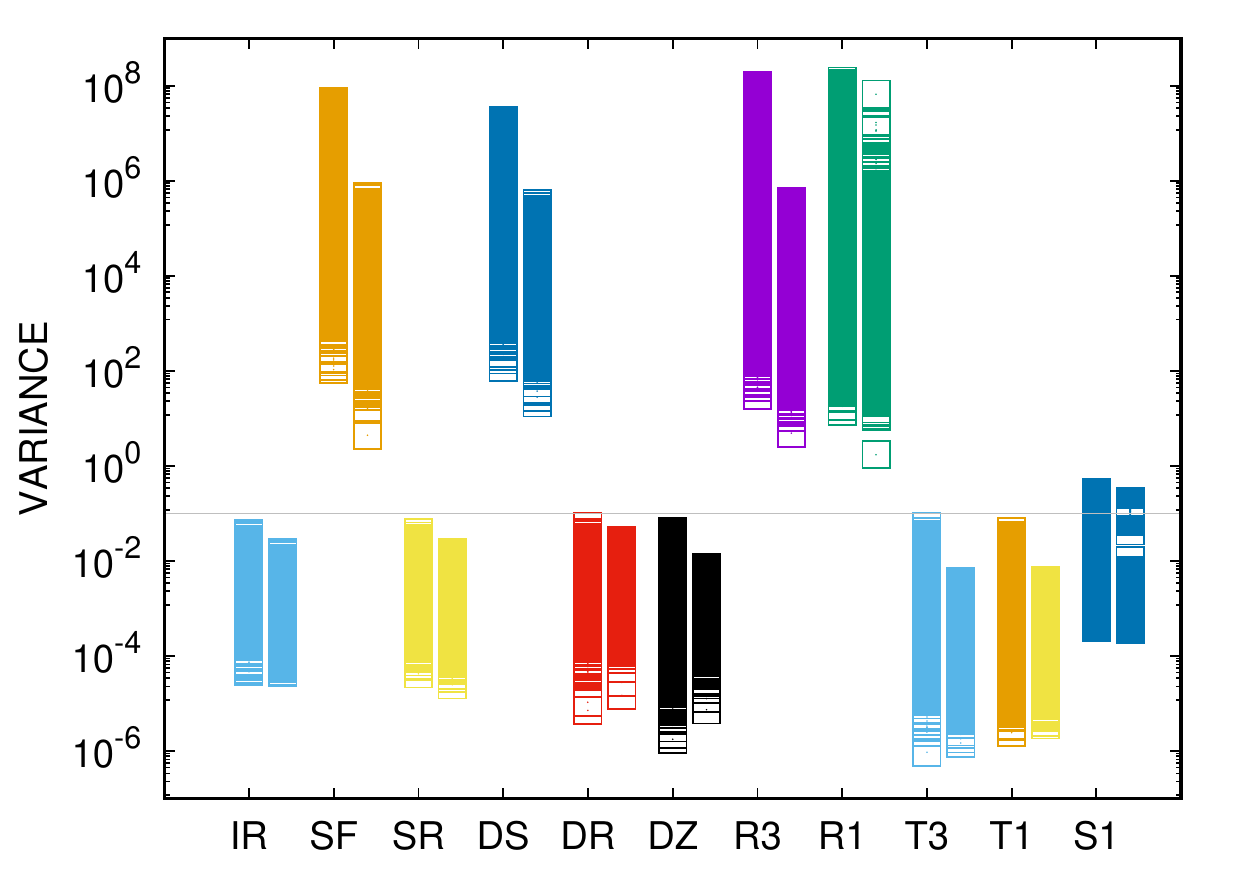}}

\caption{
Variance of models in response to 1\% Gaussian input noise: (a) stress, (b) flow, on-diagonal components on left, off-diagonal components on right.
}
\label{fig:noise_sensitivity}
\end{figure}

As a prelude to studying the dynamic stability of our plasticity TBNN model, we test the TBNN formulations' sensitivity to noise by randomly perturbing the inputs by 1\% using the CP training data.
The response to the perturbations is fairly uniform over the range strains (not shown).
As \fref{fig:noise_sensitivity} shows, the output variance for most of the models is on-par with the input variance, the exceptions being tied to the presence of the noisy invariant $\tr \sigmab$.
Clearly, pruning ill-conditioned invariants is crucial for stability.

\subsection{Prediction}

Generally speaking, errors in the predictions of the proposed TBNN plasticity models come from: errors in the elastic model, those in the flow rule, and those engendered by the integration scheme.
In preliminary work, we integrated the rate given by training data to tune the tolerances of the integration scheme and ensure the integrator error is negligible.

\begin{figure}[h!]
\centering
\subfloat[][true stress, ML flow: $\Fb_p(0)$ ensemble]
{\includegraphics[width=\figwidthtwo]{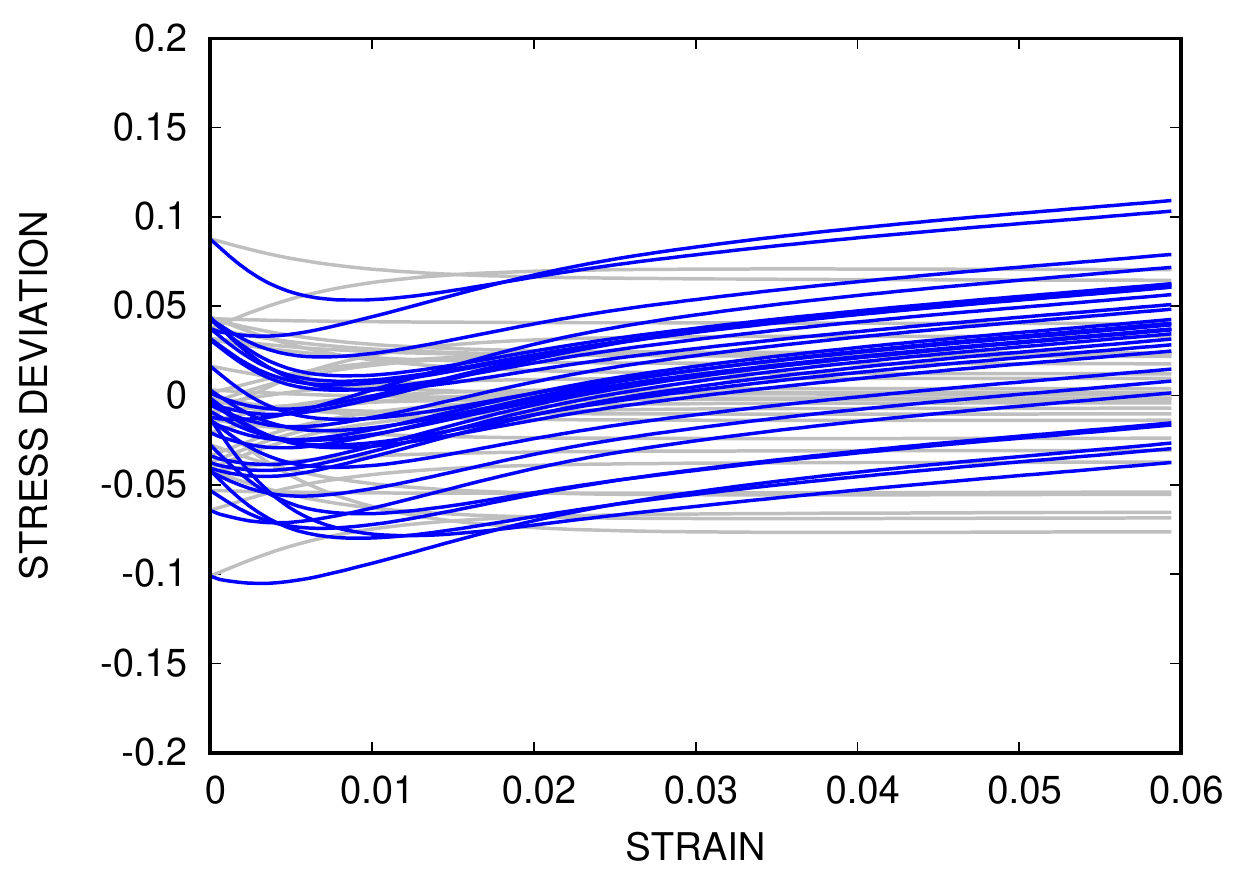}}
\subfloat[][imperfect stress, ML flow: $E$ ensemble]
{\includegraphics[width=\figwidthtwo]{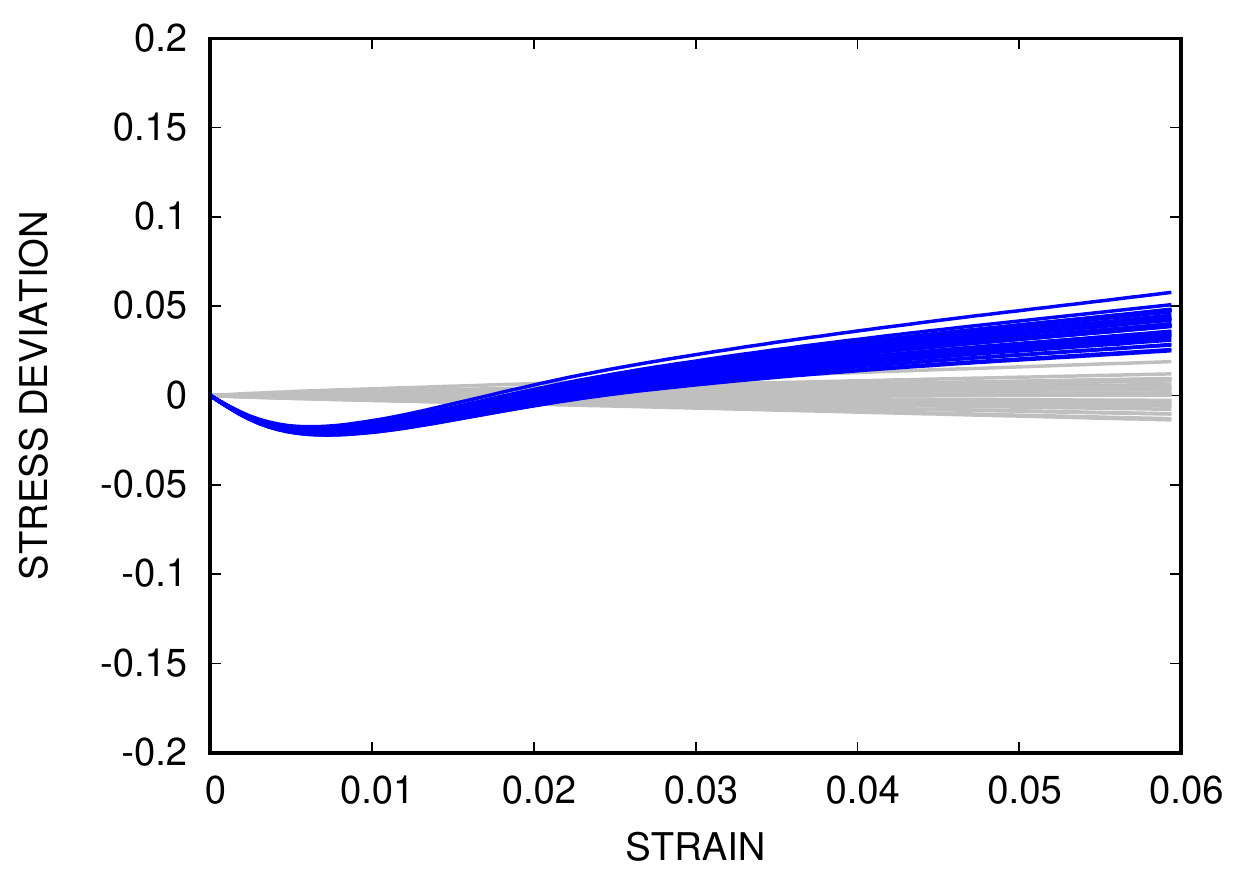}}

\subfloat[][ML stress, ML flow: $\Fb_p(0)$ ensemble]
{\includegraphics[width=\figwidthtwo]{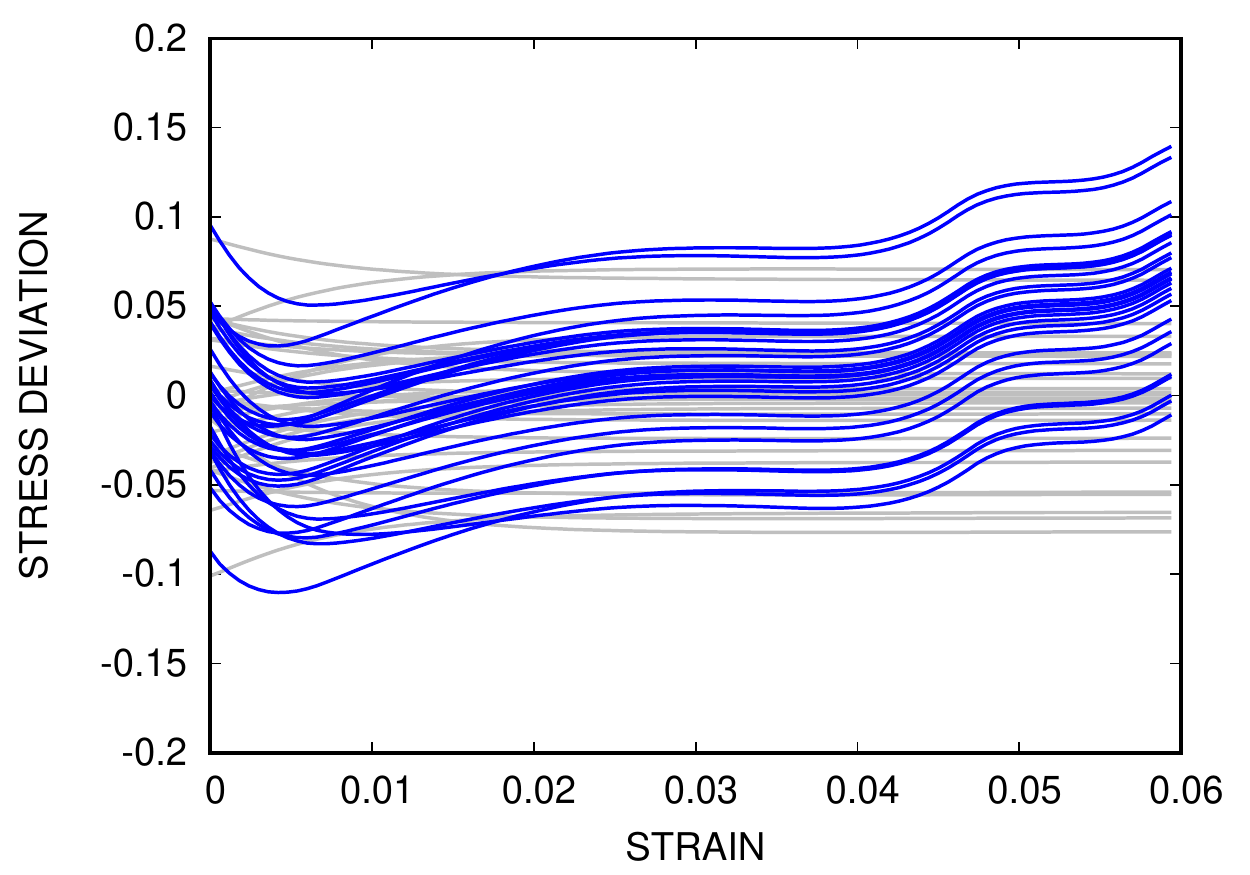}}
\subfloat[][ML stress, ML flow: model ensemble]
{\includegraphics[width=\figwidthtwo]{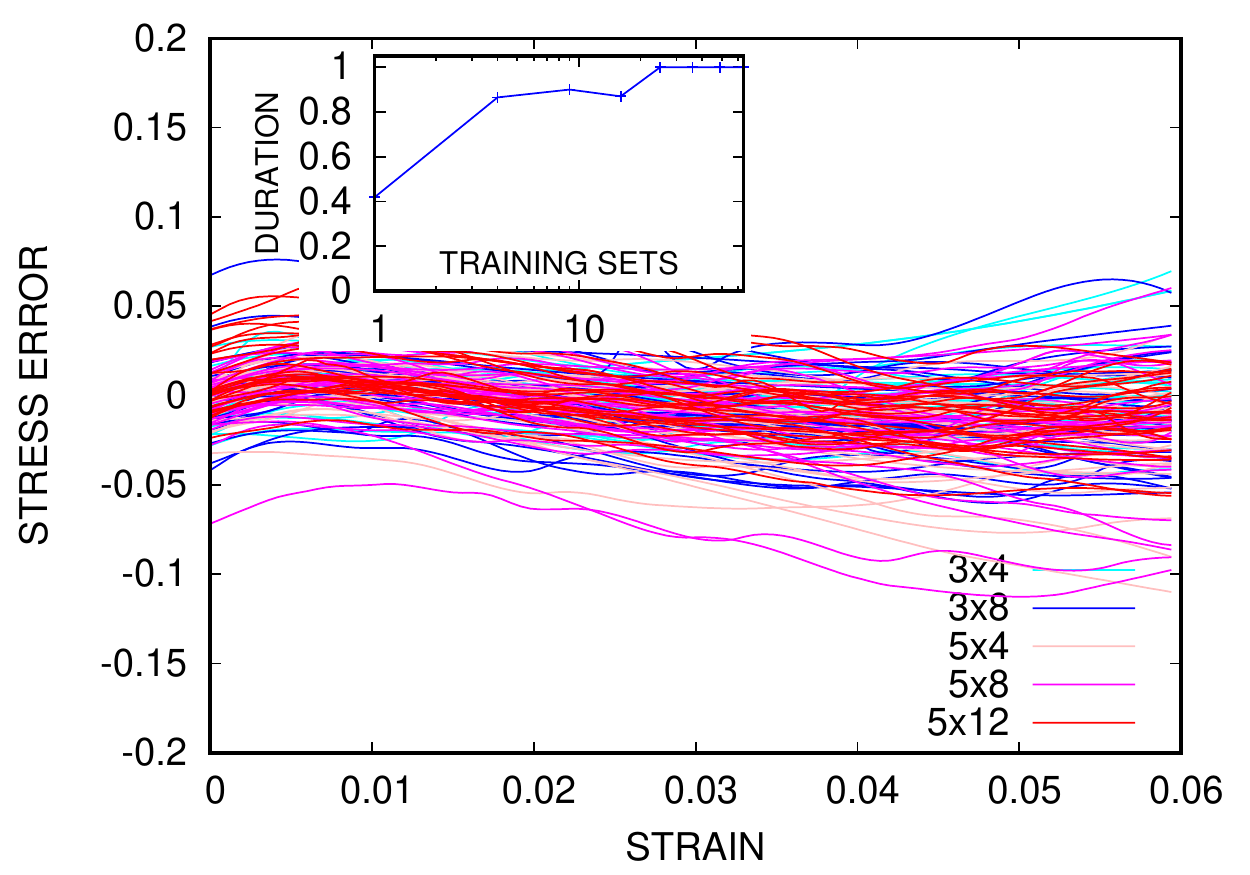}}

\caption{
Lyapunov bundle of trajectories for models on VP (known model) data:
(a) a perfect stress model and a NN flow model with perturbed initial conditions,
(b) imperfect stress models (modulus $E$ random) and a NN flow model,
(c) a NN stress model and a NN flow model with perturbed initial conditions.
(d) ensemble of (unperturbed) NN stress and NN flow models.
Deviation is with respect to an unperturbed trajectory,
gray lines: exact model, colored lines: TBNN.
Inset of (d) shows the fraction of the models that reach the double duration of the training data as a function of the amount of training data to illustrate the models' stability in extrapolation.
}
\label{fig:predict_known}
\end{figure}

In \fref{fig:predict_known} we show Lyapunov-like stability tests using a E3(3$\times$4)/T1(5$\times$8) TBNN model trained on a $D_{64}$ dataset from the known closed form VP model.
First, we perturb the initial conditions of state $\Fb_p(0)$ for a random (monotomic) loading mode $\Fb(t)$ and compare the response of the underlying model (gray lines) to that of an analytic stress (\eref{eq:known_stress})/TBNN flow model hybrid (colored) for this ensemble of initial conditions.
The TBNN response is on par with that of the true model, albeit with a distinct bias toward higher stress.
Second, we compare the same models with a $\Fb_p(0) = \Ib$ initial condition but with an imperfect stress model enacted by perturbing the Youngs' modulus $E$ of the analytic stress model.
Here again, the TBNN response is on par with the true model and yet artifacts in the trajectories are clearly present.
Third, we repeated the first investigation with at TBNN with both a ML flow and a ML stress model. 
The results are largely similar to the response of the TBNN with the true stress model albeit with additional artifacts in trajectories.
Lastly, we explore the sensitivity of the trajectory errors to flow model network size.
\fref{fig:predict_known}d shows that the trajectory errors for the flow model trained on VP data are relatively insensitive to the NN dimensions.
Also, since the variance of the results does not increase with strain, the errors are apparently primarily due to the stress representation.
The inset of \fref{fig:predict_known}d demonstrates the necessity of sufficient variety of training data. 
Here we plot the fraction of the models that reach double the training strain stably. 
Apparently, in this application, training on at least 25 trajectories is necessary to achieve robust predictions outside the training data.

\begin{figure}[h!]
\centering
\subfloat[][tension]
{\includegraphics[width=\figwidthtwo]{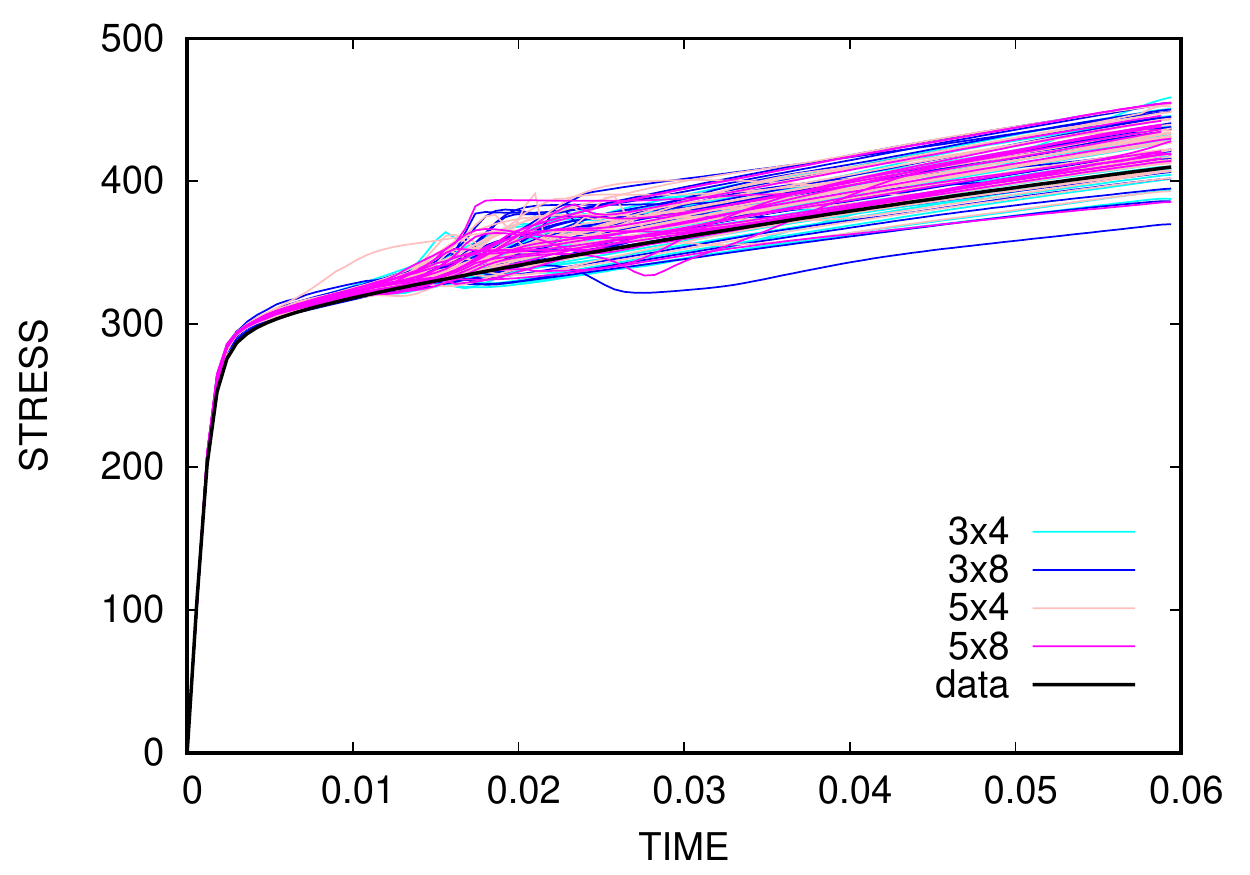}}
\subfloat[][shear]
{\includegraphics[width=\figwidthtwo]{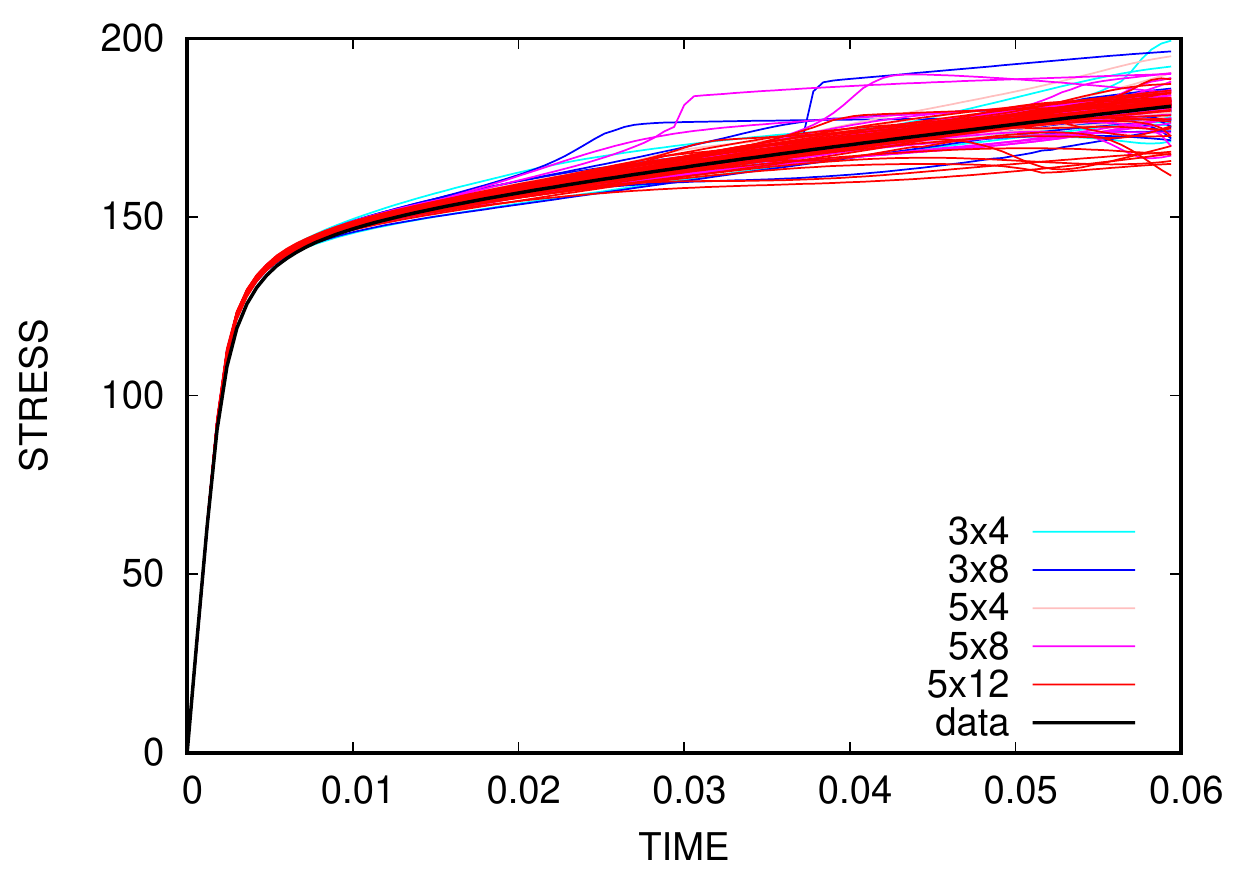}}

\subfloat[][shear+tension]
{\includegraphics[width=\figwidthtwo]{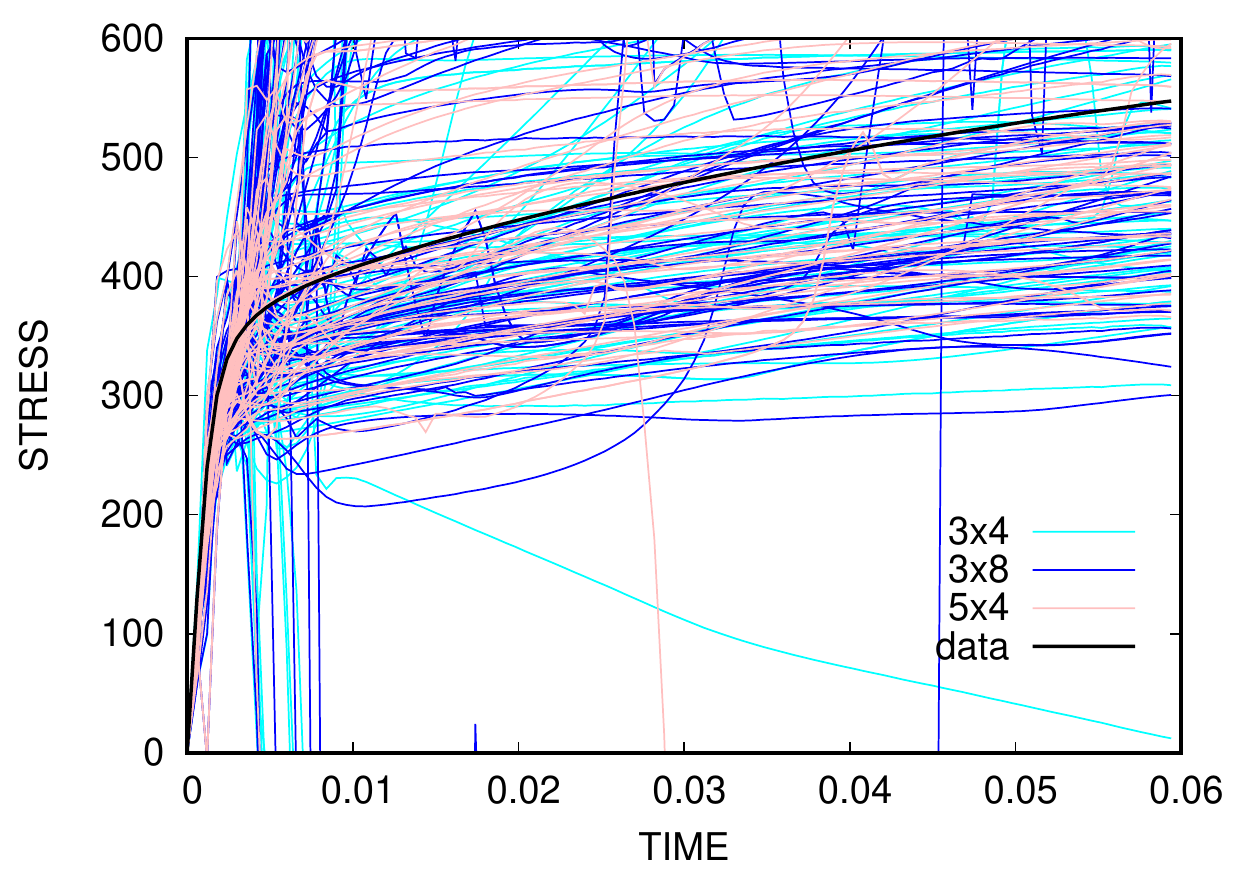}}
\subfloat[][tension,compression]
{\includegraphics[width=\figwidthtwo]{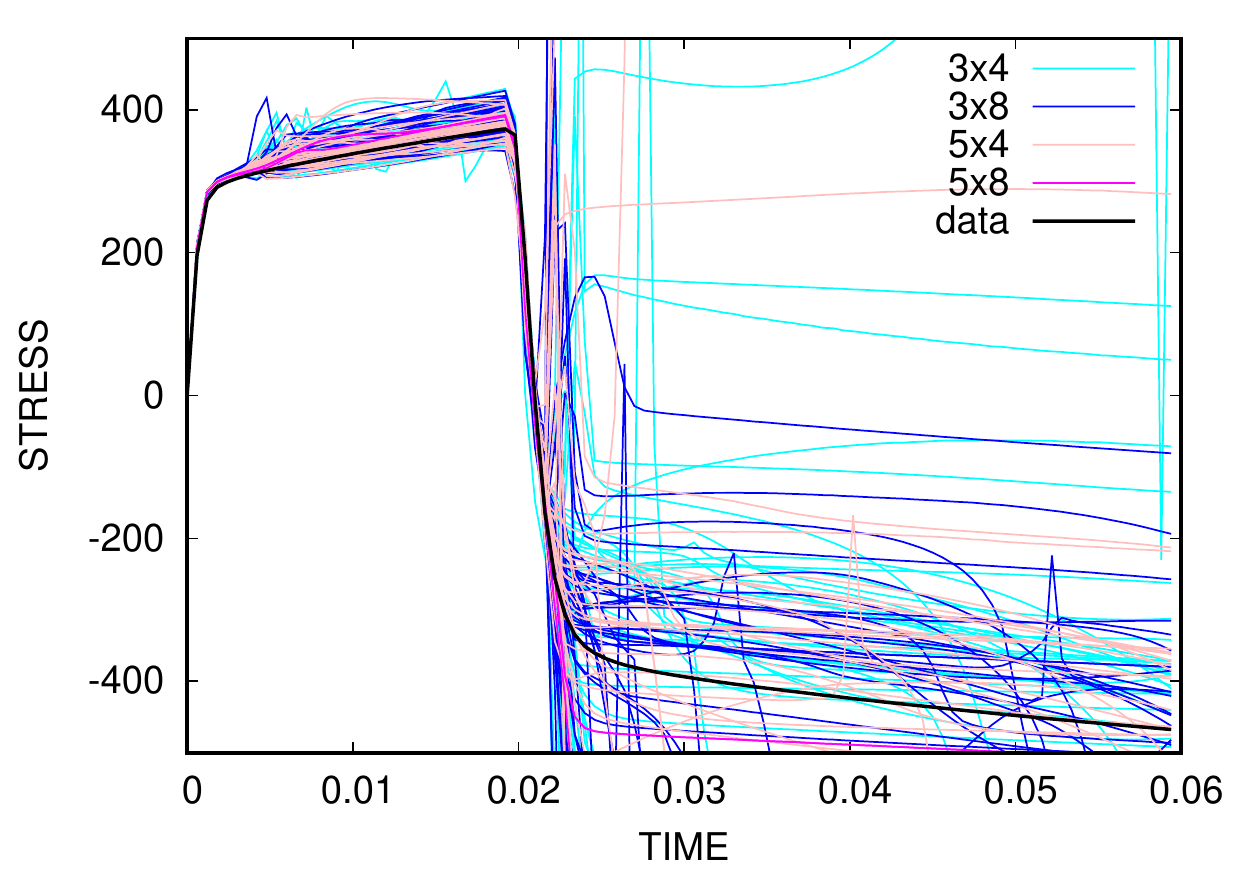}}

\caption{
Prediction of TBNN with ML stress and ML flow models trained on  CP tension and shear data:
(a) tension,
(b) shear,
(c) combined tension and shear,
(d) tension and then compression.
Note loading in (d) is at a different rate from training data.
}
\label{fig:predict}
\end{figure}

Lastly, we return to models trained on the tension and shear CP data.
\fref{fig:predict} shows the predictions of a TBNN 3$\times$4 E3 stress model with NN flow models of various sizes.
\fref{fig:predict}a,b demonstrate that the predictions are essentially self consistent with the training data. 
Also the fanning out of the trajectories is generally consistent with accumulation of errors from integrating an imperfect model.
It appears that, as the plastic flow develops, non-smooth transitions occur which make some trajectories jump to paths neighboring the true/training path.
\fref{fig:predict}c,d show the results for bona fide predictions: (c) illustrates a combined simple shear and tension loading mode, $[\Fb(t)]_{11} = (1+t)$, $[\Fb(t)]_{21} = 1/2 t$, and the other directions have traction-free boundary conditions; and (d) illustrates a non-monotonic tension-then-compression mode at a different rate, $[\Fb(t)]_{11} = (1+3t)$ for $t\in [0,0.02]$ and $[\Fb(t)]_{11} = (1.06-3t)$ for $t\in [0.02,0.06]$.
For these modes the results are considerably less stable, especially in the mixed tension-shear mode which points to the stress model being the main issue (as discussed in the previous section). 
Even the tension phase of the non-monotonic loading leads to decreased stability and accuracy compared to the tension only case, apparently due to the change in strain rate.
Lastly, in these modes none of the larger 5$\times$12 network flow models tested were stable, which gives more evidence that the main issue is a lack of sufficient variety in the training data.

\section{Discussion} \label{sec:discussion}

In this work we generalized the TBNN framework to fully take advantage of classical representation theory.
By embedding constraints and properties directly in the structure and formulation of the NNs for stress and plastic flow, we were able to reduce the amount of training required for valid models compared to current component-based NN models.
The constraints of plasticity phenomenology and the trade-offs between learning and embedding the properties lead to a variety of models and hence to a model selection process.
We showed that traditional cross-validation errors are not sufficient for the down-selection process and, for example, stability with respect to perturbation needs to be considered for a viable model.
We also illustrated the facts that: given limited data, the formulations are insensitive to a number of meta-parameters and variables, in particular the selected functional inputs; and, there are trade-offs between model complexity and property preservation, such as preserving zero-stress-at-zero-strain.
Using a known underlying data model, we demonstrated that the enhanced TBNN framework can provide robust and accurate predictions given sufficient data.
Lastly, we demonstrated that the tension (and shear) experiments traditionally used in model calibration are likely insufficient to fully train a NN model, as formulated in the TBNN framework or via a component formulation that generally displayed worse performance.

In future work we will develop an implicit time integrator based on derivatives of the neural network and explore means of obtaining sufficient variety of training data from experiments, for example using digital image correlation to obtain full-field data.

\section*{Acknowledgments}
We relied on Albany \cite{albany}, Dream3d \cite{dream3d}, Lasagne \cite{lasagne} and Theano \cite{theano} to accomplish this work. 
We also wish to thank J. Ling for helping to upgrade the TBNN package to support the modeling approach described in this paper.
This work was supported by the LDRD program at Sandia National Laboratories, and its support is gratefully acknowledged.
Sandia National Laboratories is a multimission laboratory managed and operated by National Technology and Engineering Solutions of Sandia, LLC., a wholly owned subsidiary of Honeywell International, Inc., for the U.S. Department of Energy's National Nuclear Security Administration under contract DE-NA0003525.
The views expressed in the article do not necessarily represent the views of the U.S. Department of Energy or the United States Government.




\begin{thebibliography}{10}
\expandafter\ifx\csname url\endcsname\relax
  \def\url#1{\texttt{#1}}\fi
\expandafter\ifx\csname urlprefix\endcsname\relax\def\urlprefix{URL }\fi
\expandafter\ifx\csname href\endcsname\relax
  \def\href#1#2{#2} \def\path#1{#1}\fi

\bibitem{spencer1958finite}
A.~J.~M. Spencer, R.~Rivlin, Finite integrity bases for five or fewer symmetric
  3$\times$ 3 matrices, Archive for rational mechanics and analysis 2~(1)
  (1958) 435--446.

\bibitem{spencer1958theory}
A.~J.~M. Spencer, R.~S. Rivlin, The theory of matrix polynomials and its
  application to the mechanics of isotropic continua, Archive for rational
  mechanics and analysis 2~(1) (1958) 309--336.

\bibitem{spencer1962isotropic}
A.~Spencer, R.~Rivlin, Isotropic integrity bases for vectors and second-order
  tensors, Archive for rational mechanics and analysis 9~(1) (1962) 45--63.

\bibitem{pipkin1963material}
A.~Pipkin, A.~Wineman, Material symmetry restrictions on non-polynomial
  constitutive equations, Archive for Rational Mechanics and Analysis 12~(1)
  (1963) 420--426.

\bibitem{wineman1964material}
A.~S. Wineman, A.~Pipkin, Material symmetry restrictions on constitutive
  equations, Archive for Rational Mechanics and Analysis 17~(3) (1964)
  184--214.

\bibitem{smith1964integrity}
G.~Smith, R.~Rivlin, Integrity bases for vectors--the crystal classes, Archive
  for Rational Mechanics and Analysis 15~(3) (1964) 169--221.

\bibitem{Smith1965}
G.~Smith, On isotropic integrity bases, Archive for rational mechanics and
  analysis 18~(4) (1965) 282--292.

\bibitem{rivlin1969orthogonal}
R.~Rivlin, G.~Smith, Orthogonal integrity basis for {N} symmetric matrices,
  Contributions to mechanics: Markus Reiner eightieth anniversary volume (1969)
  121.

\bibitem{spencer1971part}
A.~Spencer, Part {III}. {T}heory of invariants, Continuum physics 1 (1971)
  239--353.

\bibitem{spencer1987isotropic}
A.~Spencer, Isotropic polynomial invariants and tensor functions, in:
  Applications of tensor functions in solid mechanics, Springer, 1987, pp.
  141--169.

\bibitem{boehler1987representations}
J.-P. Boehler, Representations for isotropic and anisotropic non-polynomial
  tensor functions, in: Applications of tensor functions in solid mechanics,
  Springer, 1987, pp. 31--53.

\bibitem{zheng1994theory}
Q.-S. Zheng, Theory of representations for tensor functions--a unified
  invariant approach to constitutive equations, Applied Mechanics Reviews
  47~(11) (1994) 545--587.

\bibitem{truesdell2004non}
C.~Truesdell, W.~Noll, The non-linear field theories of mechanics, in: The
  non-linear field theories of mechanics, Springer, 2004, pp. 1--579.

\bibitem{gurtin1982introduction}
M.~E. Gurtin, An introduction to continuum mechanics, Vol. 158, Academic press,
  1982.

\bibitem{itskov2007tensor}
M.~Itskov, Tensor algebra and tensor analysis for engineers, Springer, 2007.

\bibitem{adeli1989perceptron}
H.~Adeli, C.~Yeh, Perceptron learning in engineering design, Computer-Aided
  Civil and Infrastructure Engineering 4~(4) (1989) 247--256.

\bibitem{hajela1991neurobiological}
P.~Hajela, L.~Berke, Neurobiological computational models in structural
  analysis and design, Computers \& Structures 41~(4) (1991) 657--667.

\bibitem{cheu1995automated}
R.~L. Cheu, S.~G. Ritchie, Automated detection of lane-blocking freeway
  incidents using artificial neural networks, Transportation Research Part C:
  Emerging Technologies 3~(6) (1995) 371--388.

\bibitem{theocaris1993neural}
P.~Theocaris, P.~Panagiotopoulos, Neural networks for computing in fracture
  mechanics. methods and prospects of applications, Computer Methods in Applied
  Mechanics and Engineering 106~(1-2) (1993) 213--228.

\bibitem{adeli2001neural}
H.~Adeli, Neural networks in civil engineering: 1989--2000, Computer-Aided
  Civil and Infrastructure Engineering 16~(2) (2001) 126--142.

\bibitem{ghaboussi1998autoprogressive}
J.~Ghaboussi, D.~A. Pecknold, M.~Zhang, R.~M. Haj-Ali, Autoprogressive training
  of neural network constitutive models, International Journal for Numerical
  Methods in Engineering 42~(1) (1998) 105--126.

\bibitem{furukawa1998implicit}
T.~Furukawa, G.~Yagawa, Implicit constitutive modelling for viscoplasticity
  using neural networks, International Journal for Numerical Methods in
  Engineering 43~(2) (1998) 195--219.

\bibitem{lin2008application}
Y.~Lin, J.~Zhang, J.~Zhong, Application of neural networks to predict the
  elevated temperature flow behavior of a low alloy steel, Computational
  Materials Science 43~(4) (2008) 752--758.

\bibitem{bobbili2015prediction}
R.~Bobbili, B.~Ramakrishna, V.~Madhu, A.~Gogia, Prediction of flow stress of
  7017 aluminium alloy under high strain rate compression at elevated
  temperatures, Defence Technology 11~(1) (2015) 93--98.

\bibitem{li2012comparative}
H.-Y. Li, X.-F. Wang, D.-D. Wei, J.-D. Hu, Y.-H. Li, A comparative study on
  modified zerilli--armstrong, arrhenius-type and artificial neural network
  models to predict high-temperature deformation behavior in t24 steel,
  Materials Science and Engineering: A 536 (2012) 216--222.

\bibitem{desu2014support}
R.~K. Desu, S.~C. Guntuku, B.~Aditya, A.~K. Gupta, Support vector regression
  based flow stress prediction in austenitic stainless steel 304, Procedia
  Materials Science 6 (2014) 368--375.

\bibitem{asgharzadeh2016study}
A.~Asgharzadeh, H.~J. Aval, S.~Serajzadeh, A study on flow behavior of aa5086
  over a wide range of temperatures, Journal of Materials Engineering and
  Performance 25~(3) (2016) 1076--1084.

\bibitem{ling2015evaluation}
J.~Ling, J.~Templeton, Evaluation of machine learning algorithms for prediction
  of regions of high reynolds averaged navier stokes uncertainty, Physics of
  Fluids 27~(8) (2015) 085103.

\bibitem{ling2016reynolds}
J.~Ling, A.~Kurzawski, J.~Templeton, Reynolds averaged turbulence modelling
  using deep neural networks with embedded invariance, Journal of Fluid
  Mechanics 807 (2016) 155--166.

\bibitem{tracey2015machine}
B.~D. Tracey, K.~Duraisamy, J.~J. Alonso, A machine learning strategy to assist
  turbulence model development, in: 53rd AIAA Aerospace Sciences Meeting, 2015,
  p. 1287.

\bibitem{duraisamy2015new}
K.~Duraisamy, Z.~J. Zhang, A.~P. Singh, New approaches in turbulence and
  transition modeling using data-driven techniques, in: 53rd AIAA Aerospace
  Sciences Meeting, 2015, p. 1284.

\bibitem{milano2002neural}
M.~Milano, P.~Koumoutsakos, Neural network modeling for near wall turbulent
  flow, Journal of Computational Physics 182~(1) (2002) 1--26.

\bibitem{wang2017mlturbulence}
J.-X. Wang, J.~Wu, J.~Ling, G.~Iaccarino, H.~Xiao, A comprehensive
  physics-informed machine learning framework for predictive turbulence
  modeling, arXiv preprint arXiv:1701.07102.

\bibitem{shwartz2017opening}
R.~Shwartz-Ziv, N.~Tishby, Opening the black box of deep neural networks via
  information, arXiv preprint arXiv:1703.00810.

\bibitem{koh2017understanding}
P.~W. Koh, P.~Liang, Understanding black-box predictions via influence
  functions, arXiv preprint arXiv:1703.04730.

\bibitem{alharbi2015crystal}
H.~F. Alharbi, S.~R. Kalidindi, Crystal plasticity finite element simulations
  using a database of discrete fourier transforms, International Journal of
  Plasticity 66 (2015) 71--84.

\bibitem{kirchdoerfer2016data}
T.~Kirchdoerfer, M.~Ortiz, Data-driven computational mechanics, Computer
  Methods in Applied Mechanics and Engineering 304 (2016) 81--101.

\bibitem{smith2016linking}
J.~Smith, W.~Xiong, W.~Yan, S.~Lin, P.~Cheng, O.~L. Kafka, G.~J. Wagner,
  J.~Cao, W.~K. Liu, Linking process, structure, property, and performance for
  metal-based additive manufacturing: computational approaches with
  experimental support, Computational Mechanics 57~(4) (2016) 583--610.

\bibitem{versino2017data}
D.~Versino, A.~Tonda, C.~A. Bronkhorst, Data driven modeling of plastic
  deformation, Computer Methods in Applied Mechanics and Engineering 318 (2017)
  981--1004.

\bibitem{bessa2017framework}
M.~Bessa, R.~Bostanabad, Z.~Liu, A.~Hu, D.~W. Apley, C.~Brinson, W.~Chen, W.~K.
  Liu, A framework for data-driven analysis of materials under uncertainty:
  Countering the curse of dimensionality, Computer Methods in Applied Mechanics
  and Engineering 320 (2017) 633--667.

\bibitem{shaughnessy2016efficient}
M.~Shaughnessy, R.~Jones, Efficient use of an adapting database of ab initio
  calculations to generate accurate newtonian dynamics, Journal of chemical
  theory and computation 12~(2) (2016) 664--675.

\bibitem{jain2013commentary}
A.~Jain, S.~P. Ong, G.~Hautier, W.~Chen, W.~D. Richards, S.~Dacek, S.~Cholia,
  D.~Gunter, D.~Skinner, G.~Ceder, et~al., Commentary: The materials project: A
  materials genome approach to accelerating materials innovation, {APL}
  Materials 1~(1) (2013) 011002.

\bibitem{saal2013materials}
J.~E. Saal, S.~Kirklin, M.~Aykol, B.~Meredig, C.~Wolverton, Materials design
  and discovery with high-throughput density functional theory: the open
  quantum materials database (oqmd), Jom 65~(11) (2013) 1501--1509.

\bibitem{raccuglia2016machine}
P.~Raccuglia, K.~C. Elbert, P.~D. Adler, C.~Falk, M.~B. Wenny, A.~Mollo,
  M.~Zeller, S.~A. Friedler, J.~Schrier, A.~J. Norquist,
  Machine-learning-assisted materials discovery using failed experiments,
  Nature 533~(7601) (2016) 73.

\bibitem{ling2016machine}
J.~Ling, R.~Jones, J.~Templeton, Machine learning strategies for systems with
  invariance properties, Journal of Computational Physics 318 (2016) 22--35.

\bibitem{bartok2015gaussian}
A.~P. Bart{\'o}k, G.~Cs{\'a}nyi, Gaussian approximation potentials: A brief
  tutorial introduction, International Journal of Quantum Chemistry 115~(16)
  (2015) 1051--1057.

\bibitem{khotanzad1990invariant}
A.~Khotanzad, Y.~H. Hong, Invariant image recognition by zernike moments, IEEE
  Transactions on pattern analysis and machine intelligence 12~(5) (1990)
  489--497.

\bibitem{lowe1999object}
D.~G. Lowe, Object recognition from local scale-invariant features, in:
  Computer vision, 1999. The proceedings of the seventh IEEE international
  conference on, Vol.~2, Ieee, 1999, pp. 1150--1157.

\bibitem{olver2000applications}
P.~J. Olver, Applications of Lie groups to differential equations, Vol. 107,
  Springer Science \& Business Media, 2000.

\bibitem{goodman1998representations}
R.~Goodman, N.~R. Wallach, Representations and invariants of the classical
  groups, Vol.~68, Cambridge University Press, 1998.

\bibitem{goodman2009symmetry}
R.~Goodman, N.~R. Wallach, Symmetry, representations, and invariants, Vol. 255,
  Springer, 2009.

\bibitem{sattinger2013lie}
D.~H. Sattinger, O.~L. Weaver, Lie groups and algebras with applications to
  physics, geometry, and mechanics, Vol.~61, Springer Science \& Business
  Media, 2013.

\bibitem{marsden1994mathematical}
J.~E. Marsden, T.~J. Hughes, Mathematical foundations of elasticity,
  Prentice-Hall, 1983.

\bibitem{rivlin1955further}
R.~S. Rivlin, Further remarks on the stress-deformation relations for isotropic
  materials, Journal of Rational Mechanics and Analysis 4 (1955) 681--702.

\bibitem{rivlin1997identities}
R.~S. Rivlin, G.~F. Smith, On identities for 3$\times$ 3 matrices, in:
  Collected Papers of RS Rivlin, Springer, 1997, pp. 1550--1558.

\bibitem{rivlin1955stress}
R.~S. Rivlin, J.~L. Ericksen, Stress-deformation relations for isotropic
  materials, Journal of Rational Mechanics and Analysis 4 (1955) 323--425.

\bibitem{wang1969general}
C.-C. Wang, On a general representation theorem for constitutive relations,
  Archive for Rational Mechanics and Analysis 33~(1) (1969) 1--25.

\bibitem{wang1970new}
C.-C. Wang, A new representation theorem for isotropic functions: An answer to
  professor gf smith's criticism of my papers on representations for isotropic
  functions, Archive for rational mechanics and analysis 36~(3) (1970)
  166--197.

\bibitem{seth1961generalized}
B.~Seth, Generalized strain measure with applications to physical problems,
  Tech. rep., Wisconsin University-Madison, Mathematics Research Center (1961).

\bibitem{hill1968constitutive}
R.~Hill, On constitutive inequalities for simple materials—i, Journal of the
  Mechanics and Physics of Solids 16~(4) (1968) 229--242.

\bibitem{doyle1956nonlinear}
T.~Doyle, J.~L. Ericksen, Nonlinear elasticity, in: Advances in applied
  mechanics, Vol.~4, Elsevier, 1956, pp. 53--115.

\bibitem{johnson1984discussion}
G.~C. Johnson, D.~J. Bammann, A discussion of stress rates in finite
  deformation problems, International Journal of Solids and Structures 20~(8)
  (1984) 725--737.

\bibitem{szabo1989comparison}
L.~SzABo, M.~Balla, Comparison of some stress rates, International journal of
  solids and structures 25~(3) (1989) 279--297.

\bibitem{haupt1989application}
P.~Haupt, C.~Tsakmakis, On the application of dual variables in continuum
  mechanics, Continuum Mechanics and Thermodynamics 1~(3) (1989) 165--196.

\bibitem{haupt1996stress}
P.~Haupt, C.~Tsakmakis, Stress tensors associated with deformation tensors via
  duality, Archives of Mechanics 48~(2) (1996) 347--384.

\bibitem{smith1957stress}
G.~Smith, R.~S. Rivlin, Stress-deformation relations for anisotropic solids,
  Archive for Rational Mechanics and Analysis 1~(1) (1957) 107--112.

\bibitem{smith1957anisotropic}
G.~Smith, R.~S. Rivlin, The anisotropic tensors, Quarterly of Applied
  Mathematics 15~(3) (1957) 308--314.

\bibitem{spencer1982formulation}
A.~Spencer, The formulation of constitutive equation for anisotropic solids,
  in: Mechanical Behavior of Anisotropic Solids/Comportment M{\'e}chanique des
  Solides Anisotropes, Springer, 1982, pp. 3--26.

\bibitem{zhang1990structural}
J.~Zhang, J.~Rychlewski, Structural tensors for anisotropic solids, Archives of
  Mechanics 42~(3) (1990) 267--277.

\bibitem{svendsen1994representation}
B.~Svendsen, On the representation of constitutive relations using structure
  tensors, International journal of engineering science 32~(12) (1994)
  1889--1892.

\bibitem{lee1969elastic}
E.~H. Lee, Elastic-plastic deformation at finite strains, Journal of applied
  mechanics 36~(1) (1969) 1--6.

\bibitem{lubarda2004constitutive}
V.~A. Lubarda, Constitutive theories based on the multiplicative decomposition
  of deformation gradient: Thermoelasticity, elastoplasticity, and
  biomechanics, Applied Mechanics Reviews 57~(2) (2004) 95--108.

\bibitem{Lubliner2008}
J.~Lubliner, Plasticity theory, Dover, 2008.

\bibitem{coleman1963thermodynamics}
B.~D. Coleman, W.~Noll, The thermodynamics of elastic materials with heat
  conduction and viscosity, Archive for Rational Mechanics and Analysis 13~(1)
  (1963) 167--178.

\bibitem{Simo1998}
J.~Simo, T.~Hughes, Computational Inelasticity, Springer New York, New York,
  NY, 1998.

\bibitem{Gurtin2010}
M.~E. Gurtin, E.~Fried, L.~Anand, The mechanics and thermodynamics of continua,
  Cambridge University Press, 2010.

\bibitem{taylor1934mechanism}
G.~I. Taylor, The mechanism of plastic deformation of crystals. part i.
  theoretical, Proceedings of the Royal Society of London. Series A 145~(855)
  (1934) 362--387.

\bibitem{kroner1961plastic}
E.~Kroner, On the plastic deformation of polycrystals, Acta Metallurgica 9~(2)
  (1961) 155--161.

\bibitem{bishop1951xlvi}
J.~Bishop, R.~Hill, Xlvi. a theory of the plastic distortion of a
  polycrystalline aggregate under combined stresses., The London, Edinburgh,
  and Dublin Philosophical Magazine and Journal of Science 42~(327) (1951)
  414--427.

\bibitem{bishop1951cxxviii}
J.~Bishop, R.~Hill, Cxxviii. a theoretical derivation of the plastic properties
  of a polycrystalline face-centred metal, The London, Edinburgh, and Dublin
  Philosophical Magazine and Journal of Science 42~(334) (1951) 1298--1307.

\bibitem{mandel1965generalisation}
J.~Mandel, G{\'e}n{\'e}ralisation de la th{\'e}orie de plasticit{\'e} de {WT}
  {K}oiter, International Journal of Solids and structures 1~(3) (1965)
  273--295.

\bibitem{dawson2000computational}
P.~R. Dawson, Computational crystal plasticity, International journal of solids
  and structures 37~(1-2) (2000) 115--130.

\bibitem{roters2010overview}
F.~Roters, P.~Eisenlohr, L.~Hantcherli, D.~D. Tjahjanto, T.~R. Bieler,
  D.~Raabe, Overview of constitutive laws, kinematics, homogenization and
  multiscale methods in crystal plasticity finite-element modeling: Theory,
  experiments, applications, Acta Materialia 58~(4) (2010) 1152--1211.

\bibitem{albany}
Albany: a {T}rilinos-based {PDE} code,
  \url{https://github.com/gahansen/Albany}, accessed: 2017-09-30.

\bibitem{dream3d}
Dream3d: Open, extensible software environment to allow integrated processing,
  characterization and manipulation of microstructure digitally.,
  \url{http://dream3d.bluequartz.net}, accessed: 2017-09-30.

\bibitem{clevert2015fast}
D.-A. Clevert, T.~Unterthiner, S.~Hochreiter, Fast and accurate deep network
  learning by exponential linear units (elus), arXiv preprint arXiv:1511.07289.

\bibitem{werbos1974beyond}
P.~Werbos, Beyond regression: New tools for prediction and analysis in the
  behavior science, Unpublished Doctoral Dissertation, Harvard University.

\bibitem{rumelhart1986learning}
D.~E. Rumelhart, G.~E. Hinton, R.~J. Williams, Learning representations by
  back-propagating errors, Nature 323~(6088) (1986) 533--538.

\bibitem{nielsen2015neural}
M.~A. Nielsen, Neural networks and deep learning, Determination Press, 2015.

\bibitem{sloan2004extremal}
I.~H. Sloan, R.~S. Womersley, Extremal systems of points and numerical
  integration on the sphere, Advances in Computational Mathematics 21~(1-2)
  (2004) 107--125.

\bibitem{lasagne}
Lasagne: a lightweight library to build and train neural networks in theano,
  \url{https://lasagne.readthedocs.io/en/latest/}, accessed: 2017-09-30.

\bibitem{theano}
Theano: define, optimize, and evaluate mathematical expressions involving
  multi-dimensional arrays efficiently,
  \url{http://deeplearning.net/software/theano/}, accessed: 2017-09-30.

\end{thebibliography}
\end{document}